\documentclass[prb,twocolumn,floatfix]{revtex4}
\usepackage{graphicx}

\begin{document}

\title{Numerical Renormalization Group for Impurity Quantum Phase Transitions:
Structure of Critical Fixed Points}
\author{Hyun-Jung Lee$^*$, Ralf Bulla$^*$,
and Matthias Vojta$^\dagger$}
\affiliation{$^*$Theoretische Physik III, Elektronische Korrelationen und
Magnetismus, Institut f\"ur Physik, Universit\"at Augsburg,  D-86135 Augsburg,
Germany}
\affiliation{$^\dagger$Institut f\"ur Theorie der Kondensierten Materie,
Universit\"at Karlsruhe, D-76128 Karlsruhe,
Germany}
\date{\today}

\begin{abstract}
The numerical renormalization group method is used to investigate zero
temperature phase transitions in quantum impurity systems,
in particular in the particle-hole symmetric soft-gap Anderson model.
The model displays two stable phases whose fixed points can be built up of
non-interacting single-particle states.
In contrast, the quantum phase transitions turn out to
be described by interacting fixed points, and their excitations cannot be described
in terms of free particles.
We show that the structure of the many-body spectrum of these critical fixed points
can be understood using renormalized perturbation theory close to certain values of the
bath exponents which play the role of critical dimensions.
Contact is made with perturbative renormalization group calculations for
the soft-gap Anderson and Kondo models.
A complete description of the quantum critical many-particle spectra is achieved
using suitable marginal operators;
technically this can be understood as epsilon-expansion for full many-body spectra.
\end{abstract}
\pacs{PACS:}
\vspace*{0.7cm}

\maketitle


\section{Introduction}

Zero-temperature phase transitions in quantum impurity models have recently
attracted considerable interest (for reviews see Refs.~\onlinecite{BV,MV,affleck1}).
These transitions can be observed in systems where
a zero-dimensional boundary with internal degrees of freedom (the impurity)
interacts with an extended bath of fermions or bosons.
Examples of impurity models with non-trivial phase transitions include
extensions of the Kondo model where one or two magnetic impurities couple to
fermionic baths \cite{BV}, the spin-boson model describing a two-level system
coupling to a dissipative environment \cite{Leggett,BTV}, as well as
so-called Bose-Fermi Kondo models for localized spins interacting with both
fermionic and bosonic baths.
Impurity phase transitions are of relevance for impurities in correlated
bulk systems (e.g. superconductors \cite{KV}),
for multilevel impurities like Fullerene molecules \cite{leo},
as well as for nanodevices like coupled
quantum dots \cite{marcus} or point contacts under the influence of dissipative
noise \cite{karyn, dima}.
In addition, impurity phase transitions have been argued to describe aspects
of so-called local quantum criticality in correlated lattice systems.
Here, the framework of dynamical mean-field theory is employed to map, e.g.,
the Kondo lattice model onto a single-impurity Bose-Fermi Kondo model
supplemented by self-consistency conditions, for details see Ref.~\onlinecite{edmft}.

Diverse techniques have been used to investigate impurity phase transitions,
ranging from static and dynamic large-$N$ calculations \cite{withoff},
conformal field theory \cite{affleck1},
perturbative renormalization group (RG) \cite{BV,KV,MVLF} and
the local-moment approach \cite{David,GL} to various numerical methods.
In particular, significant progress has been made using the
numerical renormalization group (NRG) technique,
originally developed by Wilson for the Kondo problem \cite{Wil75}.
The NRG combines numerically exact diagonalization with the
idea of the renormalization group, where progressively smaller energy scales
are treated in the course of the calculation.
NRG calculations are non-perturbative and are able to access arbitrarily small
energies and temperatures.
Apart from static and dynamic observables, the NRG provides information
about the many-body excitation spectrum of the system at every stage of
the RG flow. Thus, it allows to identify fixed points through their
fingerprints in the level structure.
A detailed understanding of the NRG levels is usually possible if the
fixed point can be described by non-interacting bosons or fermions -- this is
the case for most stable fixed points of impurity models, e.g., the strong-coupling
(screened) fixed point of a standard Kondo model.
Intermediate-coupling fixed points, usually being interacting, have a completely
different NRG level structure, i.e., smaller degeneracies and non-equidistant levels.
They cannot be cast into a free-particle description, with the remarkable exception
of the two-channel Kondo fixed point which is known to have a representation
in terms of free Majorana fermions \cite{BHZ}.
In general, the NRG fixed-point spectrum at impurity transitions is fully universal,
apart from a non-universal overall prefactor and discretization effects.

The purpose of this paper is to demonstrate that a complete understanding
of the NRG many-body spectrum of critical fixed points is actually possible,
by utilizing renormalized perturbation theory around a non-interacting fixed
point.
In the soft-gap Anderson model, this approach can be employed near certain
values of the bath exponent which can be identified as critical dimensions.
Using the knowledge from perturbative RG calculations, which yield the
relevant coupling constants being parametrically small near the critical dimension,
we can construct the entire quantum critical many-body spectrum from a
free-fermion model supplemented by a small perturbation.
In other words, we shall perform epsilon-expansions to determine
a complete many-body spectrum (instead of certain renormalized
couplings or observables).
Vice versa, our method allows to identify relevant degrees of freedom
and their marginal couplings by carefully analyzing the NRG spectra near
critical dimensions of impurity quantum phase transitions.

The paper is organized as follows. In Sec.~\ref{sec:sgam}
we give a brief introduction to the physics of the soft-gap
Anderson model and its quantum phase transitions.
Sec.~\ref{sec:pert} summarizes the recent results from
perturbative RG for both the soft-gap Anderson and Kondo models.
Section \ref{sec:nrg} describes the numerical
renormalization group (NRG) approach which is used here to
obtain information about the structure of the quantum
critical points. The main part of the paper is
Sec.~\ref{sec:qcp-structure} in which we discuss (i) the numerical
data for the structure of the quantum critical points
and (ii) the analytical description of these interacting
fixed points close to the upper (lower) critical dimension
$r=0$ ($r=1/2$). The main conclusions of the paper are summarized
in Sec.~\ref{sec:conclusions} where we also mention other
problems for which an analysis of the type presented here
might be useful.


\section{Soft-Gap Anderson Model}
\label{sec:sgam}

The Hamiltonian of the soft-gap Anderson model \cite{withoff}
is given by:
\begin{eqnarray}
     H &=&   \varepsilon_{f} \sum_{\sigma} f^\dagger_{\sigma} f_{\sigma}
          + U f^\dagger_{\uparrow} f_{\uparrow}
                              f^\dagger_{\downarrow} f_{\downarrow}
   \nonumber  \\
      & & + \sum_{k\sigma} \varepsilon_k c^\dagger_{k\sigma} c_{k\sigma}
       + V \sum_{k\sigma} \big( f^\dagger_{\sigma} c_{k\sigma} +
               c^\dagger_{k\sigma}  f_{\sigma} \big)  \ .
\label{eq:model}
\end{eqnarray}
This model describes the coupling of electronic degrees of freedom
at an impurity site (operators $f^{(\dagger)}_{\sigma}$)
to a fermionic bath (operators $c^{(\dagger)}_{k\sigma}$)
via a hybridization $V$. The
$f$-electrons are subject to a local Coulomb repulsion $U$, while
the fermionic bath consists of a non-interacting conduction band
with dispersion $\varepsilon_k$.
The model eq.~(\ref{eq:model}) has the same form as the
single-impurity Anderson model \cite{Hewson}
but for the soft-gap model we require that
the hybridization function
$\widetilde{\Delta}(\omega)=\pi V^2\sum_k \delta(\omega-\varepsilon_k)$ has
a soft-gap at the Fermi level,
$\widetilde{\Delta}(\omega)=\Delta \vert\omega\vert^r$,
with an exponent $r>0$.
This translates into a local conduction band density of states
$\rho(\omega) = \rho_0 |\omega|^r$ at low energies.
In the numerical calculations we used a band where this power law extends over the whole
bandwidth $D$, i.e., from $\omega=-D/2$ to $+D/2$, and we have
$\rho_0 = (2/D)^{r+1} (r+1)/2$.
However, the universal low-temperature physics to be discussed in the following
does not depend on the details of the density of states at high energies,
and consequently we will use the low-energy prefactor of the density of states, $\rho_0$,
to represent the dimensionful energy scale of the problem.
Assuming a particle-hole symmetric band, the model (\ref{eq:model}) is
particle-hole symmetric for $\epsilon_f = - U/2$.

\begin{figure}[!t]
\centerline{\includegraphics[width=0.4\textwidth]{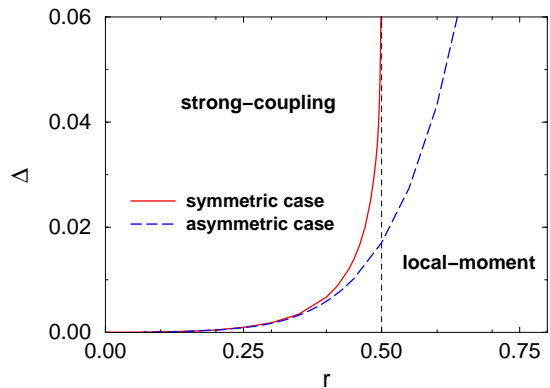}}
\caption{
      $T=0$ phase diagram for the soft-gap Anderson model
            in the p-h symmetric case (solid line, $U=10^{-3}$,
            $\varepsilon_f = -0.5 \cdot 10^{-3}$, conduction band
            cutoff at -1 and 1) and the p-h asymmetric case
            (dashed line,  $\varepsilon_f = -0.4 \cdot 10^{-3}$);
            $\Delta$ measures the hybridization strength,
            $\widetilde{\Delta}(\omega) =  \Delta |\omega|^r$.
\vspace*{-0.3cm}
}
\label{fig:pd}
\end{figure}

The soft-gap Anderson model (\ref{eq:model}) with $0<r<\infty$ displays a very rich behaviour,
in particular a continuous transition
between a local-moment (LM) and a strong-coupling (SC) phase.
Figure \ref{fig:pd} shows a typical phase diagram for the soft-gap
Anderson model.
In the particle-hole (p-h) symmetric case
(solid line) the critical coupling $\Delta_{\rm c}$
diverges at $r=\frac{1}{2}$, and no screening occurs for
$r>\frac{1}{2}$ (Refs.~\onlinecite{GBI,bullapg}).
No divergence occurs for p-h asymmetry (dashed line) \cite{GBI}.

We now briefly describe the properties of the fixed points
in the soft-gap Anderson and Kondo models \cite{GBI}.
Due to the power-law conduction band density of states, already the stable LM
and SC fixed points show non-trivial behaviour \cite{GBI,bullapg}.
The LM phase has the properties of a free spin $\frac{1}{2}$
with residual entropy $S_{\rm imp}=k_B \ln 2$ and
low-temperature impurity susceptibility $\chi_{\rm imp}=1/(4 k_B T)$,
but the leading corrections show $r$-dependent power laws.
The p-h symmetric SC fixed point has very unusual properties,
namely $S_{\rm imp}=2 r k_B \ln 2$, $\chi_{\rm imp}=r/(8 k_B T)$
for $0<r<\frac{1}{2}$.
In contrast, the p-h asymmetric SC fixed point simply displays
a completely screened moment, $S_{\rm imp}= T\chi_{\rm imp}=0$.
The impurity spectral function follows a $\omega^r$ power law
at both the LM and the asymmetric SC fixed point, whereas it
diverges as $\omega^{-r}$ at the symmetric SC fixed point --
this ``peak'' can be viewed as a generalization of the Kondo resonance in
the standard case ($r=0$), and scaling of this peak is observed upon
approaching the SC-LM phase boundary \cite{bullapg,David}.
At the critical point, non-trivial behaviour corresponding to a fractional moment
can be observed:
$S_{\rm imp}= k_B {\cal C}_S(r)$, $\chi_{\rm imp}= {\cal C}_\chi(r)/(k_B T)$
with ${\cal C}_S$, ${\cal C}_\chi$ being universal
functions of $r$ (see Refs.~\onlinecite{GBI,MVRB}).
The spectral functions at the quantum critical points
display an $\omega^{-r}$ power law (for $r<1$) with
a remarkable ``pinning'' of the critical exponent.


\section{Results from perturbative RG}
\label{sec:pert}

The Anderson model (\ref{eq:model}) is equivalent to a Kondo model when
charge fluctuations on the impurity site are negligible.
The Hamiltonian for the soft-gap Kondo model can be written as
\begin{eqnarray}
     H &=&   J \vec{S} \cdot \vec{s_0} +  \sum_{k\sigma} \varepsilon_k c^\dagger_{k\sigma} c_{k\sigma}
\label{eq:kondo}
\end{eqnarray}
where ${\vec s}(0) = \sum_{kk'\sigma\sigma'} c^\dagger_{k\sigma} {\vec \sigma}_{\sigma\sigma'}
c_{k'\sigma'} / 2$
is the conduction electron spin at the impurity site ${\bf r}\!=\!0$, and
the conduction electron density of states follows a power law
$\rho(\omega) = \rho_0 |\omega|^r$ as above.

\subsection{RG near $r=0$}

For small values of the density of states
exponent $r$, the phase transition in the
pseudogap Kondo model can be accessed from the weak-coupling limit,
using a generalization of Anderson's poor man's scaling.
Power counting about the local moment fixed point (LM) shows that
${\rm dim}[J] = -r$, i.e.,
the Kondo coupling is marginal for $r=0$.
We introduce a renormalized dimensionless Kondo coupling $j$ according to
\begin{equation}
\rho_0 J = \mu^{-r} j
\label{jren}
\end{equation}
where $\mu$ plays the role of a UV cutoff.
The flow of the renormalized Kondo coupling $j$ is given by the beta function
\begin{equation}
\label{betaj}
\beta(j) = r j - j^2 + {\cal O}(j^3)\,.
\end{equation}
For $r>0$ there is a stable fixed point at $j^\ast = 0$
corresponding to the local-moment phase (LM).
An unstable fixed point, controlling the transition
to the strong-coupling phase, exists at
\begin{equation}
\label{jast}
j^\ast = r \,,
\end{equation}
and the critical properties can be determined in a double expansion in
$r$ and $j$ \cite{KV}.
P-h asymmetry is irrelevant, i.e., a potential scattering
term $E$ scales to zero according to $\beta(e) = r e$
(where $\rho_0 E = \mu^{-r} e$),
thus the above expansion captures the p-h symmetric
critical fixed point (SCR).
As the dynamical exponent $\nu$, $1/\nu = r + {\cal O}(r^2)$, diverges as
$r\to 0^+$, $r=0$ plays the role of a lower-critical dimension of the
transition under consideration.

\subsection{RG near $r=1/2$}

For $r$ near $1/2$ the p-h symmetric critical fixed point moves
to strong Kondo coupling,
and the language of the p-h symmetric Anderson model
becomes more appropriate \cite{MVLF}.
First, the conduction electrons are integrated out exactly, yielding
a self-energy $\Sigma_f = V^2 G_{c0}$ for the $f$ electrons,
where $G_{c0}$ is the bare conduction electron Green's function
at the impurity location.
In the low-energy limit the $f$ electron propagator is then given by
\begin{equation}
G_f(i\omega_n)^{-1} = i\omega_n - i A_0\,{\rm sgn}(\omega_n)\,|\omega_n|^r
\label{dressedf}
\end{equation}
where the $|\omega_n|^r$ self-energy term dominates for $r<1$,
and the prefactor $A_0$ is
\begin{equation}
\label{A0}
A_0 = \frac{\pi \rho_0 V^2}{\cos\frac{\pi r}{2}} \,.
\end{equation}
Equation (\ref{dressedf}) describes the physics of a non-interacting
resonant level model with a soft-gap density of states.
Interestingly, the impurity spin is not fully screened for $r>0$, and
the residual entropy is $2r\ln 2$.
This precisely corresponds to the symmetric strong-coupling (SC) phase of
the soft-gap Anderson and Kondo model \cite{GBI}.

Dimensional analysis, using ${\rm dim}[f] = (1-r)/2$
[where $f$ represents the dressed fermion according to eq.~(\ref{dressedf})],
now shows that the interaction term $U$ of the Anderson model scales as
${\rm dim}[U] = 2r-1$, i.e., it is marginal at $r=1/2$.
This suggests a perturbative expansion in $U$ around the SC fixed point.
We introduce a dimensionless renormalized on-site interaction $u$ via
\begin{equation}
U = \mu^{2r-1} A_0^2 u \,.
\label{uren}
\end{equation}
The beta funcion receives the lowest non-trivial contribution
at two-loop order and reads \cite{MVLF}
\begin{eqnarray}
\label{betau}
&&\beta(u) = (1-2r)\, u -  \frac{3(\pi-2 \ln 4)}{\pi^2}\, u^3 + {\cal O}(u^5) \,.
\end{eqnarray}
For $r<1/2$ a non-interacting stable fixed point is at $u^\ast=0$ -- this is
the symmetric strong-coupling fixed point, it becomes unstable for $r>1/2$.
Additionally, for $r<1/2$ there is a pair of critical fixed points (SCR, SCR') located at
${u^\ast}^2 = \pi^2 (1-2r) / [3(\pi-2\ln 4)]$, i.e.,
\begin{equation}
\label{uast}
u^\ast = \pm 4.22 \sqrt{1/2 -r} \,.
\end{equation}
These fixed points describe the transition between an unscreened (spin or charge) moment phase and
the symmetric strong-coupling phase \cite{MVLF}.

Summarizing, both (\ref{betaj}) and (\ref{betau}) capture the same critical SCR
fixed point.
This fixed point can be accessed either by an expansion around the weak-coupling LM fixed
point, i.e., around the decoupled impurity limit, valid for $r\ll 1$, or
by an expansion around the strong-coupling SC fixed point,
i.e., around a non-interacting resonant-level (or Anderson) impurity,
and this expansion is valid for $1/2-r \ll 1$.


\section{Numerical Renormalization Group}
\label{sec:nrg}

Here we describe the numerical renormalization group method, suitably extended
to handle non-constant couplings $\widetilde\Delta(\omega)$
(see Refs.~\onlinecite{GBI,bullapg,CJ}). This method
allows a non-perturbative calculation of the many-particle spectrum and
physical properties in the whole parameter regime of the model
eq.~(\ref{eq:model}), in particular in the low-temperature limit,
so that the structure of the quantum critical points is accessible,
as discussed in Sec.~\ref{sec:qcp-structure}.

A detailed discussion of how the NRG can be applied to the soft-gap
Anderson model can be found in Refs.~\onlinecite{GBI,CJ,bullapg}.
Here we focus on those aspects of the approach necessary to
understand how information on the fixed points can be extracted.

The NRG is based on a logarithmic discretization of the
energy axis, i.e.\  one introduces a parameter $\Lambda$ and divides the
energy axis into intervals
$[-\Lambda^{-n},-\Lambda^{-(n+1)}]$ and
$[\Lambda^{-(n+1)}, \Lambda^{-n}]$ for
$n=0,1, ...., \infty$ (see Refs.~\onlinecite{Wil75,Kri80}). With some further
manipulations the original model can be mapped
onto a semi-infinite chain with the Hamiltonian
\begin{eqnarray}
  &H& =  \varepsilon_{ f} \sum_{\sigma}  f^\dagger_{\sigma}
                             f_{\sigma}
                 + U  f^\dagger_{ \uparrow} f_{ \uparrow}
                       f^\dagger_{ \downarrow} f_{ \downarrow}
                \nonumber \\
           & +& \sqrt{\frac{\xi_0}{\pi}}\sum_{\sigma} \left[
            f^\dagger_{\sigma}c_{0\sigma} +
            c^\dagger_{0\sigma}f_{\sigma}  \right]  \nonumber \\
           & +& \sum_{\sigma n=0}^\infty \left[
               \varepsilon_n c_{n\sigma}^\dagger c_{n\sigma}
               + t_n \left( c_{n\sigma}^\dagger c_{n+1\sigma}
                  + c_{n+1\sigma}^\dagger c_{n\sigma}\right)\right] \ ,
\nonumber \\
\label{eq:H_si}
\end{eqnarray}
with
\begin{equation}
   \xi_0
       = \int_{-1}^1 {\rm d} \omega\, \widetilde{\Delta}(\omega) \ .
\end{equation}
For a p-h symmetric conduction band, all the on-site energies
$\varepsilon_n$ vanish. If, in addition, the power law in
$\widetilde{\Delta}(\omega)$ extends up to a hard cut-off
$\omega_c$ (we set $\omega_c=1$), an exact expression for the
hopping matrix elements $t_n$ can be given\cite{bullapg}:
\begin{eqnarray}
    t_n &=&  \Lambda^{-n/2} \,\frac{r+1}{r+2} \,
    \frac{1-\Lambda^{-(r+2)}}{1-\Lambda^{-(r+1)}}
      \left[ 1 - \Lambda^{-(n+r+1)} \right]\nonumber \\
    &\times&
       \left[ 1 - \Lambda^{-(2n+r+1)} \right]^{-1/2}
       \left[ 1 - \Lambda^{-(2n+r+3)} \right]^{-1/2}   \label{eq:tneven}
\end{eqnarray}
  for even $n$ and
\begin{eqnarray}
    t_n &=&
  \Lambda^{-(n+r)/2} \,\frac{r+1}{r+2} \,
      \frac{1-\Lambda^{-(r+2)}}{1-\Lambda^{-(r+1)}}
      \left[ 1 - \Lambda^{-(n+1)} \right]\nonumber \\
   &\times&
       \left[ 1 - \Lambda^{-(2n+r+1)} \right]^{-1/2}
       \left[ 1 - \Lambda^{-(2n+r+3)} \right]^{-1/2} \label{eq:tnodd}
\end{eqnarray}
for odd $n$.
The semi-infinite chain is solved iteratively by
starting from the impurity and successively adding chain sites.
As the coupling $t_n$ between two adjacent sites $n$ and $n+1$ decreases as
$\Lambda^{-n/2}$ for large $n$, the low-energy states of the chain
with $n+1$ sites are generally determined by a comparatively small number
$N_{\rm s}$ of states close to the ground state of the $n$-site
system. In practice one retains only these $N_{\rm s}$ states
from the $n$-site chain to set up the Hilbert space for the $n+1$ site
chain, thus preventing the usual exponential growth of the Hilbert space
as $n$ increases. Eventually, after $n_{\rm {NRG}}$ sites have been
included in the calculation, addition of another site will not change
significantly the spectrum of many-particle excitations; the spectrum is
very close to that of the fixed point, and the calculation may be terminated.

\begin{figure}[!t]
\centerline{\includegraphics[width=0.42\textwidth]{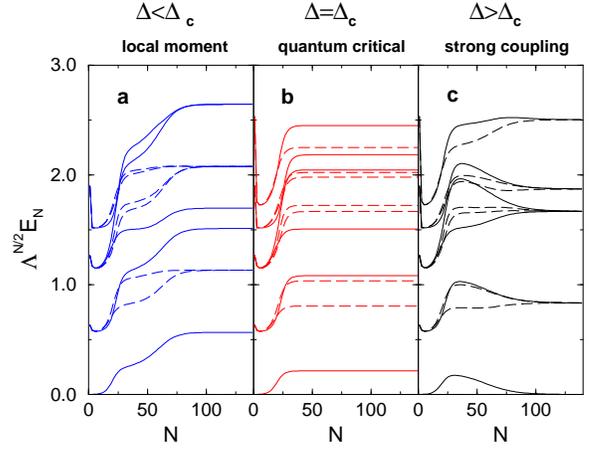}}
\caption{
 Flow diagrams for the low-energy many-body excitations
obtained from the numerical renormalization group for the three
different fixed points of the p-h symmetric soft-gap Anderson model
(exponent $r=0.4$). $N$ is the
number of iterations of the NRG procedure, $\Lambda$ the NRG discretization
parameter. Solid lines: $(Q,S)=(1,0)$, dashed lines: $(Q,S)=(0,1/2)$.
\vspace*{-0.3cm}
}
\label{fig:flow}
\end{figure}

In this way, the NRG iteration gives the many-particle energies $E_N$
for a sequence of Hamiltonians $H_N$ which correspond to the
Hamiltonian eq.~(\ref{eq:H_si}) by the replacement
\begin{equation}
 \sum_{\sigma n=0}^\infty \longrightarrow \sum_{\sigma n=0}^{N-1} \ .
\end{equation}
An example for the
dependence of  the lowest lying energy levels
on the chain length (the flow diagram) is given in
Fig.~\ref{fig:flow}c for the soft-gap
Anderson model with $r=0.4$, $D=2$, $U/D = 10^{-3}$ and $\Delta= 0.0075$;
the parameters used for the NRG calculations are $\Lambda=2$ and
$N_{\rm s} =300$.
The states are labelled by the quantum numbers $Q$
(which characterizes the number of particles measured relative to
half-filling), and the total spin, $S$ [solid lines in
Fig.~\ref{fig:flow} are for $(Q,S)=(1,0)$, dashed lines
for $(Q,S)=(0,1/2)$].
As mentioned above, the
energy scale is reduced in each step by a factor $\Lambda^{1/2}$.
To allow for a direct comparison of the energies for different chain lengths,
it is thus convenient to plot $\Lambda^{N/2}E_{N}$ instead of the eigenvalues
$E_{N}$ of the $N$-site chain directly.
Note that here and in the following we use the convention that the energies
shown in the flow diagrams are proportional to the bandwidth $D$.

As is apparent from
Fig.~\ref{fig:flow}c,
the properties of the system in this case do not change further for chain
lengths $n_{\rm NRG}>120$.
Without going into details now, one can show that the distribution of energy
levels for $N>120$ in Fig.~\ref{fig:flow}c
is characteristic of the SC phase of the model
(see Sec.~\ref{sec:qcp-structure}).

If by contrast we choose instead a value of $\Delta=0.006$,
we obtain the flow diagram shown in Fig.~\ref{fig:flow}a.
Here it is evident that the
fixed point level structure is entirely different from the SC solution,
and indeed this particular pattern is now characteristic of the LM phase
of the model. We can thus conclude, simply from inspection of the two flow
diagrams, that the critical $\Delta_{\rm c}$ separating the SC
and LM phases of the soft-gap Anderson model for the model parameters
specified, lies in the interval $[0.006, 0.0075]$.

Tuning the value of $\Delta$ to the critical value $\Delta_{\rm c}$,
results in the flow diagram of Fig.~\ref{fig:flow}b. Apparently,
the structure of the fixed point at $\Delta_{\rm c}$ neither
coincides with the SC nor with the LM fixed point. It is
clear that it cannot be simply constructed from single-particle
states as for the SC and LM fixed points. An important observation is that
certain degeneracies present in the SC or LM fixed points are
lifted at the QCP. As shown in the following section, a further
hint on the structure of the QCPs is given by the dependence
of their many-particle spectra on the bath exponent $r$.


\section{Structure of the Quantum Critical Points}
\label{sec:qcp-structure}

\begin{figure}[!t]
\centerline{\includegraphics[width=0.4\textwidth]{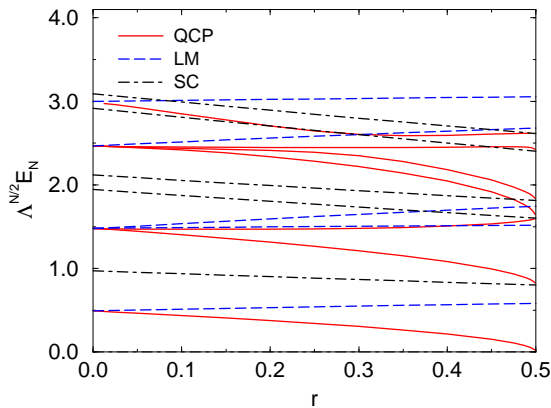}}
     \caption{Dependence of the many-particle spectra for the three
            fixed points of the p-h symmetric soft-gap Anderson model on
            the exponent $r$: SC (black dot-dashed lines),
            LM (blue dashed lines),
            and the (symmetric) quantum critical point (red solid lines).
            The data are shown for the subspace $Q=1$ and $S=0$ only.}
     \label{fig1}
  \end{figure}

In  Fig.~\ref{fig1}, the many-particle spectra
of the three fixed points
(SC: dot-dashed lines, LM: dashed lines, and QCP: solid lines)
of the symmetric soft-gap model
are plotted as functions of the exponent $r$
(for a similar figure, see Fig.~13 in Ref.~\onlinecite{GBI}).
The data are shown for an odd number of sites
only and we select the lowest lying energy levels
for the subspace $Q=1$ and $S=0$.

As usual, the fixed point structure of the strong coupling and
local moment phases can be easily constructed from the
single-particle states of a free conduction electron chain.
This is discussed in more detail later.
Let us now turn to the line of quantum critical points.
What information can be extracted from Fig.~\ref{fig1} to understand
the structure of these fixed points?

First we observe that the levels of the quantum critical points,
$E_{N,\rm QCP}(r)$, approach the levels of the LM (SC)
fixed points in the limit $r\to 0$ ($r\to 1/2$):
\begin{eqnarray}
     \lim_{r\to 0} \left\{ E_{N,\rm QCP}(r)  \right\}
           &=& \left\{ E_{N,\rm LM}(r=0)  \right\} \ , \nonumber \\
     \lim_{r\to 1/2} \left\{ E_{N,\rm QCP}(r)  \right\}
           &=& \left\{ E_{N,\rm SC}(r=1/2)  \right\} \ , 
\end{eqnarray}
where $\{ \ldots \}$ denotes the whole set of many-particle states.

\begin{figure}[!t]
\centerline{\includegraphics[width=0.45\textwidth]{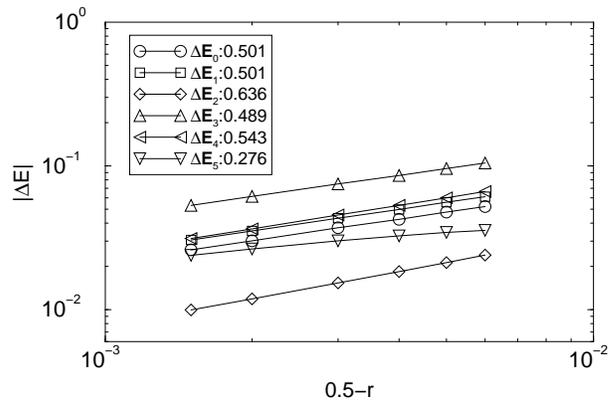}}
     \caption{Difference $\Delta E$ between the energy levels of QCP and SC
              fixed points close to $r= 1/2$ in a double-logarithmic plot.
              The inset shows the values of the exponents obtained
              from a fit to the data points.}
     \label{fig:diff}
  \end{figure}

For $r\to 0$, each individual many-particle level $E_{N,\rm QCP}(r)$ deviates
linearly from the levels of the LM fixed point, while the
deviation from the SC levels is proportional to
$\sqrt{1/2-r}$ for $r\to 1/2$. This is illustrated in Fig.~\ref{fig:diff}
where we plot a selection of energy differences $\Delta E$ between
levels of QCP and SC fixed points close to $r= 1/2$.
The inset shows the values of the exponents obtained
from a fit to the data points. For some levels, there are significant
deviations from the exponent $1/2$. This is because the
correct exponent is only obtained in the limit $r\to 1/2$
(the QCP levels have been obtained only up to $r=0.4985$).

Note that the behaviour of the
many-particle levels close to $r= 1/2$ has direct
consequences for physical properties at the QCP;
the critical exponent of the local susceptibility at the QCP, for example, shows a square-root
dependence on $(1/2-r)$ close to $r=1/2$, see Ref.~\onlinecite{GBI}.

In both limits, $r\to 0$ and $r\to 1/2$,
we observe that degeneracies due to the
combination of single-particle levels, present at the
LM and SC fixed points, are lifted at the quantum critical
fixed points as soon as one is moving away from
$r=0$ and  $r=1/2$, respectively. This already suggests that
the quantum critical point is interacting
and cannot be constructed from
non-interacting single-particle states.

In the remainder of the paper we want to show
how to connect this information from NRG to the perturbative RG
of Sec.~\ref{sec:pert}.
We know that the critical fixed point can be accessed via
two different epsilon-expansions \cite{KV,MVLF} near the two critical dimensions,
and, combined with renormalized perturbation theory,
these expansions can be used to evaluate various observables near
criticality.
Here, we will employ this concept to perform renormalized perturbation theory
for the entire many-body spectrum at the critical fixed point.
To do so, we will start from the many-body spectrum of
one of the trivial fixed points, i.e., LM near $r=0$ and
SC near $r=1/2$, and evaluate corrections to it in lowest-order
perturbation theory.
This will be done within the NRG concept working directly with the discrete many-body
spectra corresponding to a finite NRG chain (which is diagonalized numerically).
As the relevant energy scale of the spectra decreases as $\Lambda^{-n/2}$
along the NRG iteration, the strength of the perturbation has to be
scaled as well, as the goal is to capture a {\em fixed point} of
the NRG method.
This scaling of the perturbation follows precisely from its scaling
dimension -- the perturbation marginal at the value of $r$ corresponding to
the critical dimension.
With the proper scaling, the operator which we use to capture the
difference between the free-fermion and critical fixed points
becomes exactly marginal [see eqs.~(\ref{hnp1}) and (\ref{hnp2}) below].

\subsection{Perturbation theory close to $r=0$}

\begin{figure}[ht]
\centerline{\includegraphics[width=0.45\textwidth]{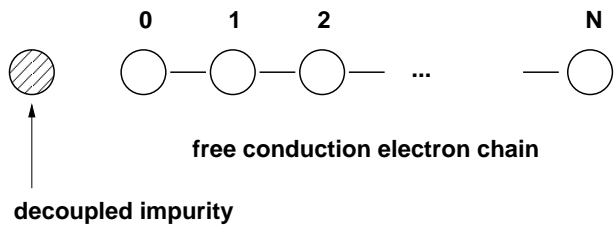}}
\vspace*{0.5cm}
   \caption{The spectrum of the LM fixed point is described
    by the impurity decoupled from the free conduction electron
    chain.}
     \label{fig:lmfp}
\end{figure}

Let us now describe in detail the analysis of the deviation
of the QCP levels from the LM levels close to $r=0$ (the
case $r= 1/2$ is discussed
in Sec.~\ref{subsec:B}).
An effective description of the LM fixed point is given
by a finite chain with the impurity decoupled from the
conduction electron part, see Fig.~\ref{fig:lmfp}.

The conduction electron part of the effective Hamiltonian
is given by
\begin{equation}
   H_{{\rm c},N} = \sum_{\sigma n=0}^{N-1}
                t_n \left( c_{n\sigma}^\dagger c_{n+1\sigma}
                  + c_{n+1\sigma}^\dagger c_{n\sigma}\right) \ .
\label{eq:Hclmfp}
\end{equation}
As usual, the structure of the fixed point spectra depends
on whether the total number of sites is even or odd. To simplify
the discussion in the following, we only consider a total {\em
odd} number of sites (the flow diagrams of Fig.~\ref{fig:flow} are all calculated
for this case). For the LM fixed point, this means that the number
of sites, $N+1$, of the free conduction electron chain is even,
so $N$ in eq.~(\ref{eq:Hclmfp}) is odd.

\begin{figure}[ht]
\centerline{\includegraphics[width=0.35\textwidth]{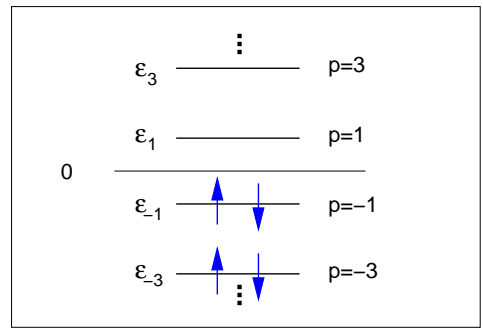}}
\vspace*{0.5cm}
   \caption{Single-particle spectrum of the free conduction electron
chain eq.~(\ref{eq:Hdiag-lm}). The ground state is given by all the levels
with $p<0$ filled.}
     \label{fig:lmfp-spl}
\end{figure}

The single-particle spectrum of the free chain with an even
number of sites, corresponding to the diagonalized Hamiltonian
\begin{equation}
   \bar{H}_{{\rm c},N} =
      \sum_{\sigma p} \epsilon_p\, \xi^\dagger_{p\sigma}\xi_{p\sigma}\ ,
\label{eq:Hdiag-lm}
\end{equation}
is sketched in Fig.~\ref{fig:lmfp-spl}.
(Note that the $\epsilon_p$ have to be evaluated numerically.)

As we assume p-h symmetry, the positions of the single-particle
levels are symmetric with respect to $0$ with
\begin{equation}
  \epsilon_p = -\epsilon_{-p} \ \ , \ \ p=1,3,\ldots,N \ ,
\end{equation}
and
\begin{equation}
 \sum_p \equiv \sum_{p=-N,\ p\ {\rm odd}}^{p=N}
  \ .
\label{eq:sum_p}
\end{equation}
Note that an equally spaced spectrum of single-particle levels
is only recovered in the limit $\Lambda\to 1$ (see Fig.~6 in
Ref.~\onlinecite{BHZ}) for the case $r=0$.

The RG analysis of Sec.~\ref{sec:pert} tells us that the critical
fixed point is perturbative accessible from the LM one
using a Kondo-type coupling as perturbation.
We thus focus on the operator
\begin{equation}
\label{hnp1}
   H_N^\prime = \alpha(r) f(N) \,
                \vec{S}_{\rm imp} \cdot \vec{s}_0 \,,
\end{equation}
with the goal to calculate the many-body spectrum of the critical
fixed point via perturbation theory in $H_N^\prime$ for small $r$.
The function $\alpha(r)$ contains the fixed-point value of the Kondo-type coupling,
and $f(N)$ will be chosen such that $H_N^\prime$ is exactly marginal,
i.e., the effect of $H_N^\prime$ on the many-particle
energies decreases as $\Lambda^{-N/2}$ which is the same
$N$ dependence which governs the scaling of the many-particle
spectrum itself.
The scaling analysis of Sec.~\ref{sec:pert}, eqs. (\ref{jren}) and (\ref{jast}),
suggests a parametrization of the coupling as
\begin{equation}
\label{alphapar}
\alpha(r) = \frac{\mu^{-r}}{\rho_0} \, \alpha\, r \,,
\end{equation}
where $\rho_0$ is the prefactor in the density of states,
and $\mu$ is a scale of order of the bandwidth --
such a factor is required here to make
$\alpha$ a {\em dimensionless} parameter.
Thus, the strength of the perturbation increases linearly
with $r$ at small $r$ (where $\mu^{-r}/\rho_0 = D + {\cal O}(r)$ for a
featureless $|\omega|^r$ density of states).

The qualitative influence of the operator
$\vec{S}_{\rm imp} \cdot \vec{s}_0$ on the many-particle states
has been discussed in general in Ref.~\onlinecite{GBI} for finite $r$ and
in Refs.~\onlinecite{Wil75,Kri80}
for $r=0$.
Whereas an antiferromagnetic exchange
coupling is marginally relevant in the gapless case ($r=0$), it turns
out to be irrelevant for finite $r$, see Ref.~\onlinecite{GBI}.
This is of course consistent with the scaling analysis
of Sec.~\ref{sec:pert}: the operator (\ref{hnp1}) simply
represents a Kondo coupling, with a tree-level scaling
dimension of ${\rm dim}[J] = -r$.
A detailed analysis of the $N$-dependence of the operator
$\vec{S}_{\rm imp} \cdot \vec{s}_0$ shows that it decreases
as
$\Lambda^{-Nr/2}\Lambda^{-N/2}=\Lambda^{-N(r+1)/2}$ with increasing $N$.
Consequently, we have to choose
\begin{equation}
\label{fn1}
   f(N) = \Lambda^{Nr/2} \, .
\end{equation}
This result also directly follows from ${\rm dim}[J] = -r$:
As the NRG discretization yields a decrease of the running energy scale
of $\Lambda^{-N/2}$, the $\vec{S}_{\rm imp} \cdot \vec{s}_0$ term in
$H_N^\prime$ (\ref{hnp1}) scales as $\Lambda^{-Nr/2}$.
The function $f(N)$ is now simply chosen to compensate this effect;
using eq.~(\ref{fn1}) the operator $H_N^\prime$ becomes exactly
marginal.

\begin{figure}[ht]
\centerline{\includegraphics[width=0.35\textwidth]{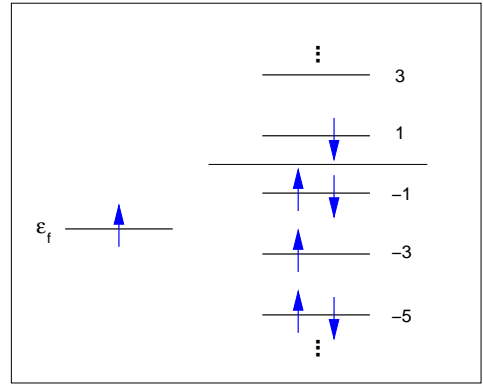}}
\vspace*{0.5cm}
   \caption{One possible excitation with energy $\epsilon_1 + \epsilon_3$
            and quantum numbers $Q=0$ and $S_z=+1/2$.}
     \label{fig:lmfp-ex}
\end{figure}

Now we turn to a discussion of the many-body spectrum.
The relevant ground state of the effective model for the LM fixed point
consists of the filled impurity level (with one electron with
either spin $\uparrow$ or $\downarrow$) and all the conduction
electron states with $p<0$ filled with both $\uparrow$ and $\downarrow$,
as shown in Fig.~\ref{fig:lmfp-spl}. Let us now focus on excitations
with energy
$\epsilon_1 + \epsilon_3$ measured with respect to the ground state.
Figure \ref{fig:lmfp-ex} shows one such
excitation; in this case, one electron with
spin $\downarrow$ is removed from the $p=-3$ level and one electron
with spin $\downarrow$ is added to the $p=1$ level. The impurity
level is assumed to be filled with an electron with spin $\uparrow$,
so the resulting state has $Q=0$ and $S_z=+1/2$. In total, there
are 32 states with excitation energy $\epsilon_1 + \epsilon_3$.
These states can be classified using the quantum numbers $Q$, $S$, and
$S_z$.

Here we consider only the states with quantum numbers
$Q=0$, $S=1/2$, and $S_z=1/2$ (with excitation energy
$\epsilon_1 + \epsilon_3$) which form a four-dimensional
subspace. As the state shown in Fig.~\ref{fig:lmfp-ex} is not an eigenstate
of the total spin $S$, we have to form proper linear combinations
to obtain a basis for this subspace; this basis can be written
in the form:
\begin{eqnarray}
  \vert \psi_1 \rangle &=& \frac{1}{\sqrt{2}} f^\dagger_{\uparrow}
        \left( \xi^\dagger_{1\uparrow} \xi_{-3\uparrow} +
               \xi^\dagger_{1\downarrow} \xi_{-3\downarrow}
        \right)
        \vert \psi_0 \rangle     \nonumber \\
  \vert \psi_2 \rangle &=& \left[
        \frac{1}{\sqrt{6}} f^\dagger_{\uparrow}
        \left( \xi^\dagger_{1\uparrow} \xi_{-3\uparrow} -
               \xi^\dagger_{1\downarrow} \xi_{-3\downarrow}
        \right) +
        \frac{2}{\sqrt{6}} f^\dagger_{\downarrow}
        \xi^\dagger_{1\uparrow} \xi_{-3\downarrow}
        \right] \vert \psi_0 \rangle     \nonumber \\
  \vert \psi_3 \rangle &=& \frac{1}{\sqrt{2}} f^\dagger_{\uparrow}
        \left( \xi^\dagger_{3\uparrow} \xi_{-1\uparrow} +
               \xi^\dagger_{3\downarrow} \xi_{-1\downarrow}
        \right)
        \vert \psi_0 \rangle     \nonumber \\
  \vert \psi_4 \rangle &=& \left[
        \frac{1}{\sqrt{6}} f^\dagger_{\uparrow}
        \left( \xi^\dagger_{3\uparrow} \xi_{-1\uparrow} -
               \xi^\dagger_{3\downarrow} \xi_{-1\downarrow}
        \right) +
        \frac{2}{\sqrt{6}} f^\dagger_{\downarrow}
        \xi^\dagger_{3\uparrow} \xi_{-1\downarrow}
        \right] \vert \psi_0 \rangle    \nonumber \\
     \label{eq:four-psis}
\end{eqnarray}
where the state $\vert \psi_0 \rangle$ is given by the product of the
ground state of the conduction electron chain and the
empty impurity level:
\begin{equation}
   \vert \psi_0 \rangle = \left[ \prod_{p<0}
          \xi^\dagger_{p\uparrow} \xi^\dagger_{p\downarrow}
          \vert 0 \rangle_{\rm cond}
          \right] \otimes \vert 0 \rangle_{\rm imp} \ .
\end{equation}

The fourfold degeneracy of the subspace ($Q=0$, $S=1/2$, $S_z=1/2$)
of the LM fixed point at energy
$\epsilon_1 + \epsilon_3$ is partially split for finite $r$ in the
spectrum of the quantum critical fixed point.
Let us now calculate the influence of the perturbation
$H_N^\prime$ on the states
$\vert \psi_1 \rangle,\ldots \vert \psi_4 \rangle$, concentrating
on the splitting of the energy levels up to first order.
Degenerate perturbation theory requires the calculation of
the matrix
\begin{equation}
   W_{ij} = \langle \psi_i\vert  H_N^\prime \vert \psi_j \rangle
   \ \ , \ \
   i,j=1,\ldots 4 \ \ ,
\end{equation}
and a subsequent calculation of the eigenvalues of
$\left\{W_{ij}\right\}$ gives the level splitting.

Details of the calculation of the matrix elements $W_{ij}$ are
given in Appendix \ref{app:A}. The result is
\begin{equation}
   \left\{W_{ij}\right\} = \alpha(r) f(N)
    \left[
      \begin{array}{cccc}
         0 & \frac{\sqrt{3}}{4}\gamma & 0 & 0 \\
         \frac{\sqrt{3}}{4}\gamma & -\frac{1}{2}\beta  & 0 & 0 \\
         0 & 0 & 0 & \frac{\sqrt{3}}{4}\gamma\\
         0 & 0 & \frac{\sqrt{3}}{4}\gamma & -\frac{1}{2}\beta
      \end{array}
    \right]  \ ,
\label{eq:Wij}
\end{equation}
with $\gamma = \left[
             \vert \alpha_{01} \vert^2 - \vert \alpha_{0-3} \vert^2
              \right] $
and $\beta =  \left[
             \vert \alpha_{01} \vert^2 + \vert \alpha_{0-3} \vert^2
              \right] $.
The $N$-dependence of the coefficients $\alpha_{0p}$ (which
relate the operators $c_{0\sigma}$ and $\xi_{p\sigma}$, see
eq.~(\ref{eq:c-xi})) is given by
\begin{equation}
   \vert \alpha_{0p} \vert^2 \propto \Lambda^{-Nr/2}\Lambda^{-N/2} \ ,
\end{equation}
(see also Sec.~III A in Ref.~\onlinecite{GBI}).
Numerically we find that
\begin{eqnarray}
    \gamma &=& -0.1478 \cdot \Lambda^{-Nr/2}\Lambda^{-N/2} \nonumber\\
    \beta &=& 2.0249 \cdot \Lambda^{-Nr/2}\Lambda^{-N/2} \ ,
\end{eqnarray}
where the prefactors depend on the exponent $r$ and the
discretization parameter $\Lambda$ (the quoted
values are for $r=0.01$ and $\Lambda=2.0$).
The matrix $\left\{W_{ij}\right\}_{r=0.01}$ then takes the form
\begin{eqnarray}
  &&\hspace*{-1.5cm}
\left\{W_{ij}\right\}_{r=0.01}  = \alpha (r=0.01)\, \Lambda^{-N/2}
      \nonumber \\
   &\times &\left[
      \begin{array}{cccc}
         0 &-0.064  & 0 & 0 \\
         -0.064 & -1.013  & 0 & 0 \\
         0 & 0 & 0 & -0.064\\
         0 & 0 & -0.064 &  -1.013
      \end{array}
    \right]  \ .
\end{eqnarray}
Diagonalization of this matrix gives the first-order corrections
to the energy levels
\begin{eqnarray}
    \Delta E_1(r=0.01)&=& \Delta E_3(r=0.01)
 \nonumber \\
         &=&\alpha (r=0.01) \, \Lambda^{-N/2}\cdot
      (-1.0615) \nonumber \\
    \Delta E_2(r=0.01) &=& \Delta E_4(r=0.01) \nonumber \\
         &=&
      \alpha(r=0.01) \, \Lambda^{-N/2}\cdot 0.0004
\end{eqnarray}
with
\begin{eqnarray}
    E_{N,\rm QCP}(r=0.01,i) &=&
         E_{N,\rm LM}(r=0.01,i)
\nonumber \\
&+& \Delta E_i(r=0.01) \ ,
\end{eqnarray}
($i=1,\ldots 4$).
Apparently, the fourfold degeneracy of the subspace ($Q=0$, $S=1/2$, $S_z=1/2$)
with energy
$\epsilon_1 + \epsilon_3$ is split in two levels which are
both twofold degenerate.

We repeated this analysis for a couple of other subspaces and a list of
the resulting matrices $\left\{W_{ij}\right\}$ and the energy shifts
$\Delta E$ is given in Appendix \ref{app:A}.

Let us now proceed with the comparison of the perturbative results
with the structure of the quantum critical fixed point calculated
from the NRG.
For our specific choice of the conduction band density of states,
the relation (\ref{alphapar}) yields $\alpha(r) = \alpha\, r \, D$
for small $r$ (where $\mu^r \approx 1$).
Using the corresponding equations for the energy shifts
in Appendix \ref{app:A}, we observe that a {\em single} parameter
$\alpha$ must be sufficient to describe the level shifts in {\em all}
subspaces, provided that the exponent $r$ is small enough so that
the perturbative calculations are still valid. A numerical
fit gives $\alpha \approx 1.03$ for  $\Lambda=2.0$,
(the $\Lambda$-dependence of
$\alpha$ is discussed later, see Fig.~\ref{fig:alpha}).

\begin{figure}[!t]

\unitlength1cm
\begin{picture}(5,6)

\put(-1.5,0){
\includegraphics[width=0.21\textwidth]{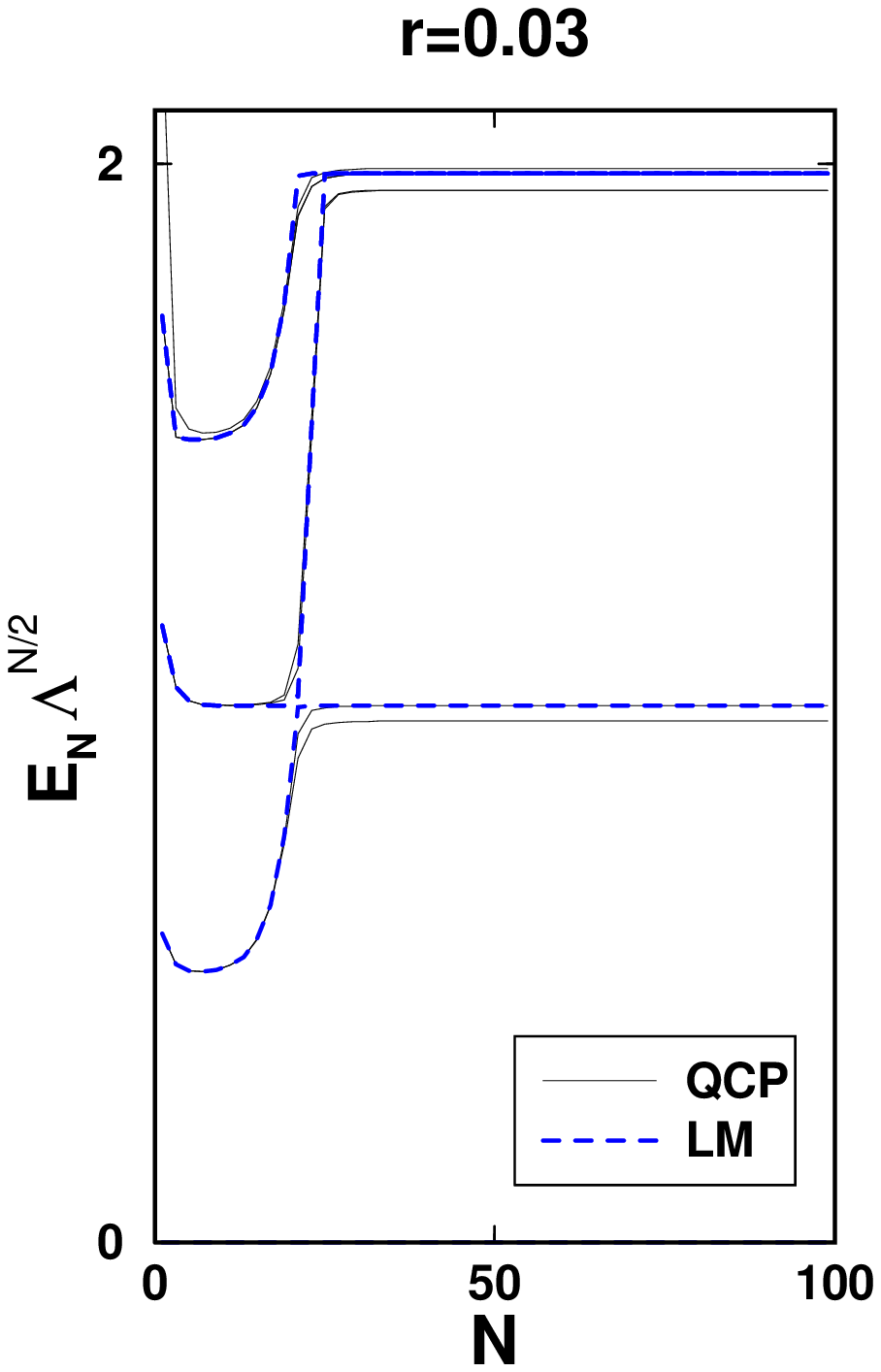}}
\put(2.5,0){
\includegraphics[width=0.21\textwidth]{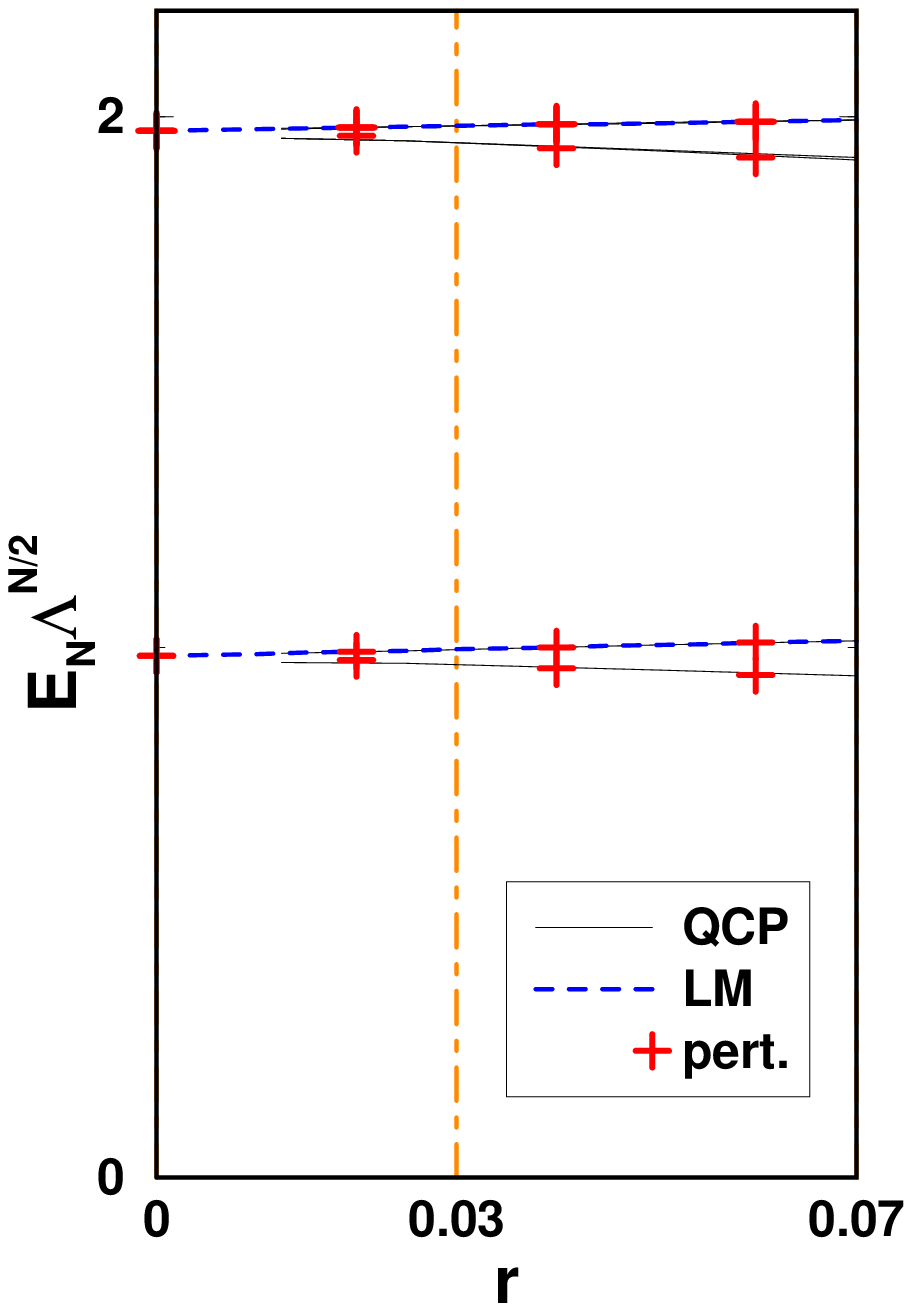}}

\put(-1.5,4){\bf a)}
\put(2.5,4){\bf b)}

\end{picture}

\caption{a) Flow diagram of the lowest lying energy levels
for $r=0.03$; dashed lines: flow to the LM fixed point;
solid lines: flow to the quantum critical fixed point. b) The deviation
of the QCP levels from the LM levels increases linearly with
$r$. This deviation together with the splitting of the
energy levels can be explained by the perturbative
calculation (crosses) as described in the text.
}
\label{fig:compare_LM}
\end{figure}

\begin{figure}[ht]
\centerline{\includegraphics[width=0.40\textwidth]{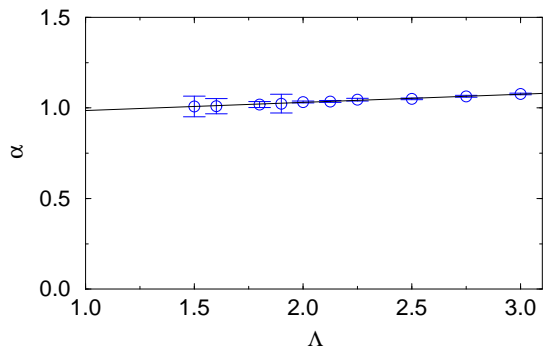}}
\vspace*{0.5cm}
   \caption{Dependence of the coupling parameter $\alpha$ on the
            NRG-discretization parameter $\Lambda$. The circles
            correspond to the NRG data and the
            solid line is a linear fit to the data:
            $\alpha(\Lambda) = 0.985 + 0.045(\Lambda - 1.0)$.}
     \label{fig:alpha}
\end{figure}

Figure \ref{fig:compare_LM} summarizes the NRG results
together with the perturbative analysis for exponents $r$
close to 0. A flow diagram of the lowest lying energy levels
 is shown in Fig.~\ref{fig:compare_LM}a for a small value of the
exponent, $r=0.03$, so that the levels of the QCP only slightly
deviate from those of the LM fixed point. As discussed above,
the deviation
of the QCP levels from the LM levels increases linearly with
$r$, see Fig. \ref{fig:compare_LM}b.
We indeed get a very good agreement between the
perturbative result (crosses) and the NRG-data (lines)
for exponents  up to $r \approx 0.07$.
The data shown here are for the subspaces
($Q=0$, $S=1/2$, $S_z=1/2$) and energy $2\epsilon_1$
(the levels at $E_N\Lambda^{N/2}\approx 1$, see
Appendix \ref{sec:sub2}) and
($Q=0$, $S=1/2$, $S_z=1/2$) and energy $\epsilon_1 + \epsilon_3$
(the levels at $E_N\Lambda^{N/2}\approx 2$, see the example discussed
in this section).

In the NRG, the continuum limit
corresponds to the limit $\Lambda\to1$, but due to the
drastically increasing numerical effort upon reducing $\Lambda$,
results for the  continuum limit have to be obtained via
extrapolation of NRG data for $\Lambda$ in,
for example, the range
$1.5 < \Lambda <  3.0$.
 Figure \ref{fig:alpha} shows the numerical
results from the NRG calculation together with
a linear fit to the data:
            $\alpha(\Lambda) = 0.985 + 0.045(\Lambda - 1.0)$.
Taking into account the increasing error bars for smaller
values of $\Lambda$, the extrapolated value
$\alpha(\Lambda\to 1)\approx 0.985$ is in excellent agreement with the
result from the perturbative RG calculation, which is directly
for the continuum limit and gives $\alpha=1.0$.

\subsection{Perturbation theory close to $r=1/2$}
\label{subsec:B}

\begin{figure}[ht]
\centerline{\includegraphics[width=0.45\textwidth]{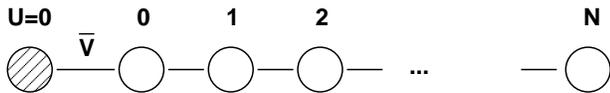}}
\vspace*{0.5cm}
   \caption{The spectrum of the SC fixed point is described
    by the non-interacting impurity coupled to the free conduction electron
    chain.}
     \label{fig:scfp}
\end{figure}

To describe the deviation of the QCP levels from the SC levels
close to $r=1/2$, we have to start from an effective description
of the SC fixed point. This is given by a finite chain
{\em including} the impurity site with the Coulomb repulsion
$U=0$ at the impurity site and a hybridization $\bar{V}$ between
impurity and the first conduction electron site,
see Fig.~\ref{fig:scfp}.

Note that the SC fixed point can also
be described by the limit $\bar{V}\to\infty$ and finite $U$
which means that impurity
and first conduction electron site are effectively 
removed from the chain. This
reduces the number of sites of the chain by two and leads to
exactly the same level structure as including the impurity
with $U=0$. However, the description with the impurity
included (and $U=0$) is
more suitable for the following analysis.

The corresponding effective Hamiltonian is that of a soft-gap
Anderson model on a finite chain with $N+2$ sites and
$\varepsilon_{f}=U=0$ (i.e., a p-h symmetric resonant level model).

\begin{equation}
  H_{{\rm sc},N} =
             \bar{V} \sum_{\sigma} \left[
            f^\dagger_{\sigma}c_{0\sigma} +
            c^\dagger_{0\sigma}f_{\sigma}  \right]
          + H_{{\rm c},N} \ ,
\end{equation}
with $H_{{\rm c},N}$ as in eq.~(\ref{eq:Hclmfp}).

\begin{figure}[ht]
\centerline{\includegraphics[width=0.35\textwidth]{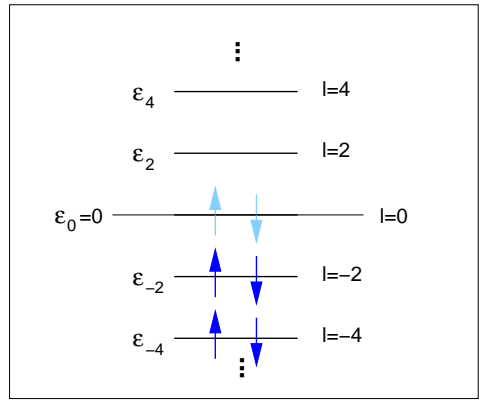}}
\vspace*{0.5cm}
   \caption{Single-particle spectrum of the free conduction electron
chain eq.~(\ref{eq:Hscfp}). The ground state is fourfold degenerate
with all the levels with $l<0$ filled and the level $l=0$
either empty, singly ($\uparrow$ or $\downarrow$) or
doubly occupied.}
     \label{fig:scfp-spl}
\end{figure}

As for the effective description of the LM fixed point, the
effective Hamiltonian is that of a free chain. Focussing,
as above, on odd values of $N$, the total number of sites
of this chain, $N+2$, is odd. The single-particle spectrum
of the free chain with an odd number of sites, corresponding to
the diagonalized Hamiltonian
\begin{equation}
   \bar{H}_{{\rm sc},N} =
      \sum_{\sigma l} \epsilon_l\, \xi^\dagger_{l\sigma}\xi_{l\sigma}\ ,
\label{eq:Hscfp}
\end{equation}
is sketched in Fig.~\ref{fig:scfp-spl}.
As we assume p-h symmetry, the positions of the single-particle
levels are symmetric with respect to $0$ with
\begin{equation}
  \epsilon_0 = 0 \ \ , \ \
  \epsilon_l = -\epsilon_{-l} \ \ , \ \ l=2,4,\ldots,(N+1) \ ,
\end{equation}
and
\begin{equation}
 \sum_l \equiv \sum_{l=-(N+1),\ l\ {\rm even}}^{l=N+1}
  \ .
\end{equation}
The ground state of the effective model for the SC fixed
point is fourfold degenerate, with all levels with $l<0$ filled
and the level $l=0$
either empty, singly ($\uparrow$ or $\downarrow$) or
doubly occupied.

According to Sec.~\ref{sec:pert} the proper perturbation
to access the critical fixed point from the SC one
is an on-site repulsion, thus we choose
\begin{equation}
   H_N^\prime = \beta(r) \bar{f}(N) \,
\left(n_{f\uparrow} - \frac{1}{2}\right)
\left(n_{f\downarrow} - \frac{1}{2}\right) ,
\label{eq:pertU}
\label{hnp2}
\end{equation}
($n_{f\sigma}=f_{\sigma}^{\dagger}f_{\sigma}$) 
with the strength of the perturbation parametrized as
\begin{equation}
        \beta(r) = \mu^{2r-1} \rho_0^2 \, \bar{V}^4 \, \beta \sqrt{1/2 - r} \ .
\end{equation}
see Sec.~\ref{sec:pert}.
Note that $\rho_0^2(r=1/2) = 9/(2D^3)$ for a featureless power-law density of
states with bandwidth $D$.
The $N$ dependence of the operator 
$\left(n_{f\uparrow} - \frac{1}{2}\right)
\left(n_{f\downarrow} - \frac{1}{2}\right)$
is given by
$\Lambda^{(r-1/2)N}\Lambda^{-N/2}=\Lambda^{(r-1)N}$, so we have to choose
\begin{equation}
\label{fn2}
    \bar{f}(N) = \Lambda^{(1/2-r)N} \ .
\end{equation}
This again follows from the scaling analysis of Sec.~\ref{sec:pert}:
the on-site repulsion has scaling dimension ${\rm dim}[U] = 2r-1$.
Thus the 
$\left(n_{f\uparrow} - \frac{1}{2}\right)
\left(n_{f\downarrow} - \frac{1}{2}\right)$
 term in
$H_N^\prime$ (\ref{hnp2}) scales as $\Lambda^{N(r-1/2)}$,
and $\bar{f}(N)$ (\ref{fn2}) compensates this behavior to make
$H_N^\prime$ exactly marginal.

We continue with analyzing the low-lying many-body levels.
Similar as above, we focus on one specific example, these
are excitations with energy $2\varepsilon_2$ measured with respect to the
ground state and quantum numbers $Q=-1$, $S=0$, and $S_z=0$.
This subspace is two-dimensional and the basis is given by
\begin{eqnarray}
  \vert \psi_1 \rangle &=&
        -  \xi^\dagger_{0\uparrow}  \xi^\dagger_{0\downarrow}
           \xi_{-2\uparrow} \xi_{-2\downarrow}
        \vert \psi_0 \rangle  \ ,   \nonumber \\
  \vert \psi_2 \rangle &=&
        \frac{1}{\sqrt{2}}
        \left( \xi^\dagger_{2\uparrow}  \xi_{-2\uparrow}
         + \xi^\dagger_{2\downarrow} \xi_{-2\downarrow}
        \right)\vert \psi_0 \rangle  \ ,
\end{eqnarray}
with
\begin{equation}
   \vert \psi_0 \rangle =  \prod_{l<0}
          \xi^\dagger_{l\uparrow} \xi^\dagger_{l\downarrow}
          \vert 0 \rangle
           \ .
\label{eq:psi0_U}
\end{equation}
(Note that in this definition of $\vert \psi_0 \rangle$,
the $l=0$-level is empty.)

The two-fold degeneracy of this subspace is lifted for
$r<1/2$ in the spectrum of the quantum critical points.
The matrix
$W_{ij} = \langle \psi_i\vert  H_N^\prime \vert \psi_j \rangle$
($i,j=1,2$) is given by
\begin{equation}
   \left\{W_{ij}\right\} = \beta(r) \bar{f}(N)|\alpha_{f2}|^4
    \left[
      \begin{array}{cc}
         2-2\kappa+{\kappa^2} & 2\sqrt{2}\kappa  \\
         2\sqrt{2}\kappa & 2+{\kappa^2}      \end{array}
    \right]  \ ,
\end{equation}
with $\kappa=|\alpha_{f0}|^2/|\alpha_{f2}|^2$.
 The {\it N}-dependence of the coefficients $|\alpha_{fl}|$
(which relate the operators $f_{\mu}$ and $\xi_{l\mu}$, see
eq.~(\ref{eq:f-xi}))  is given by
\begin{equation}
   \vert \alpha_{fl} \vert^2 \propto \Lambda^{(r-1)N/2} \ ,
\end{equation}
Numerically we find that
\begin{eqnarray}
    |\alpha_{f2}|^2 &=& 0.0366 \cdot (D/\bar{V})^{2} \Lambda^{(r-1)N/2}\nonumber\\
    |\alpha_{f0}|^2 &=& 0.0930 \cdot (D/\bar{V})^{2} \Lambda^{(r-1)N/2} \ ,
\end{eqnarray}
where the prefactors depend on the exponent $r$ and the quoted
value is for $r=0.499$.
The matrix $\left\{W_{ij}\right\}_{r=0.499}$ then takes the form
\begin{eqnarray}
   \left\{W_{ij}\right\}_{r=0.499} &=&
     \beta(r=0.499) (D/\bar{V})^{4} \Lambda^{-N/2} \nonumber \\
    &\times&    \left[
      \begin{array}{cc}
         0.0044  &0.0094 \\
         0.0094 & 0.011 \\
      \end{array}
    \right]  \ ,
\end{eqnarray}
Diagonalization of this matrix gives the first-order corrections
to the energy levels
\begin{eqnarray}
    \Delta E_1(r=0.499)&=&\beta(r=0.499) (D/\bar{V})^{4}\,\Lambda^{-N/2}
    \cdot(-0.0023) \nonumber \\
    \Delta E_2(r=0.499) &=& \beta(r=0.499) (D/\bar{V})^{4}\,\Lambda^{-N/2}
   \cdot(0.018) \nonumber \\
\end{eqnarray}
with
\begin{eqnarray}
    E_{N,\rm QCP}(r=0.499,i) &=& \nonumber \\
         E_{N,\rm SC}(r=0.499,i) &+& \Delta E_i(r=0.499) \ ,
\end{eqnarray}
($i=1, 2$).
We repeated this analysis for a couple of other subspaces and a list of
the resulting matrices $\left\{W_{ij}\right\}$ and the energy shifts
$\Delta E$ is given in Appendix \ref{app:B}.

\begin{figure}[!t]

\unitlength1cm
\begin{picture}(5,6)

\put(-1.5,0){
\includegraphics[width=0.21\textwidth]{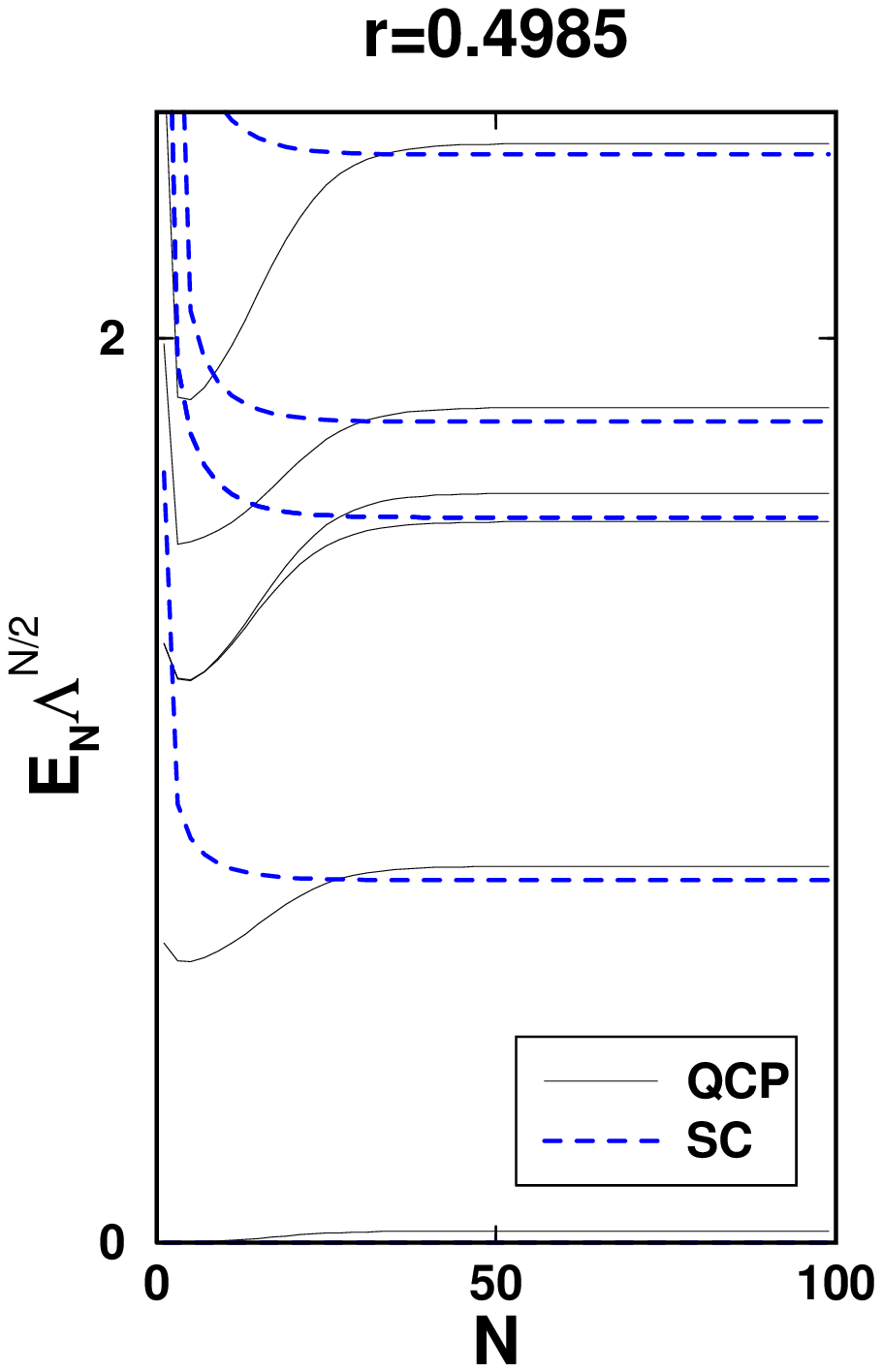}}
\put(2.5,0){
\includegraphics[width=0.21\textwidth]{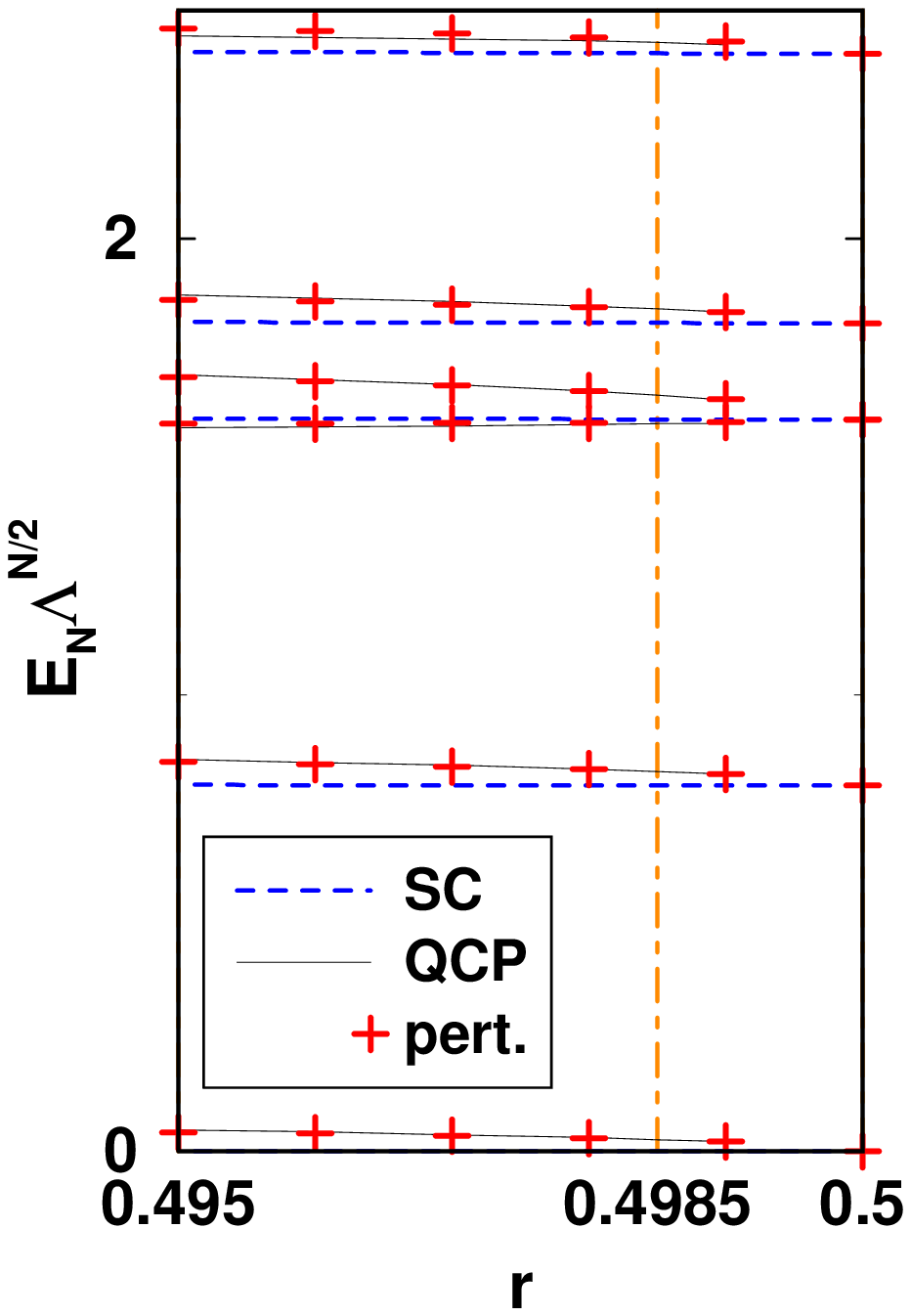}}

\put(-1.5,4){\bf a)}
\put(2.5,4){\bf b)}

\end{picture}

\caption{a) Flow diagram of the lowest lying energy levels
for $r=0.4985$; dashed lines: flow to the SC fixed point;
solid lines: flow to the quantum critical fixed point. b) The deviation
of the QCP levels from the SC levels is proportional to
$\sqrt{1/2-r}$. This deviation together with the splitting of the
energy levels can be explained by the perturbative
calculation (crosses) as described in the text.
}
\label{fig:compare_SC}
\end{figure}

The comparison of the perturbative results with the numerical results from the NRG calculation is
shown in Fig.~\ref{fig:compare_SC}b.
As for the case $r\approx 0$ we observe that a single parameter $\beta$ is sufficient
to describe the level shifts in all subspaces, provided the exponent $r$ is close enough
to $r=1/2$ so that the perturbative calculations are valid. For $\Lambda=2.0$ we
find  $\beta \approx 70$ and the $\Lambda\to1$ extrapolation results in
$\beta(\Lambda\to1) \approx 73.0\pm 5.0$ (the error bars here are significantly
larger as for the extra\-polation of the coupling $\alpha$).
The results from perturbative RG, Sec.~\ref{sec:pert}, specifically
eqs. (\ref{uren}) and (\ref{uast}),
yield $\beta(r) = \mu^{2r-1} \rho_0^2 \bar{V}^4 \, 2 \pi^2 u^\ast$.
This gives $\beta = 83.3$.

Similar to Fig.~\ref{fig:compare_LM} above, we show in  Fig.~\ref{fig:compare_SC}a a flow diagram for an exponent
very close to  $1/2$, $r=0.4985$, so that the levels of the QCP only slightly
deviate from those of the SC levels. As discussed above, this deviation
is proportional to $\sqrt{1/2-r}$, see Fig.~\ref{fig:compare_SC}b.
The data shown here are all for subspaces with
($Q=-1$, $S=0$, $S_z=0$); the unperturbed energies $E$ of
these subspaces are:
\begin{itemize}
  \item $E=0$: the levels at $E_N\Lambda^{N/2}\approx 0$, see App.~\ref{sec:subx1},
  \item $E=\epsilon_2$: the levels at $E_N\Lambda^{N/2}\approx 0.8$, see App.~\ref{sec:subx2},
  \item $E=2\epsilon_2$: the levels at $E_N\Lambda^{N/2}\approx 1.6$, see the example discussed
              in this section,
  \item $E=\epsilon_4$: the levels at $E_N\Lambda^{N/2}\approx 1.8$, see App.~\ref{sec:subx3},
  \item $E=3\epsilon_2$: the levels at $E_N\Lambda^{N/2}\approx 2.4$.
\end{itemize}
We again find a very good
agreement between the perturbative results (crosses) and the NRG data (lines).

Thus we can summarize that our renormalized perturbation
theory for the NRG many-body spectrum works well in the
vicinity of both $r=0$ and $r=1/2$.
In principle, from the many-body spectrum (and suitable matrix elements)
all other observables like thermodynamic data and dynamic correlation
functions can be calculated.
We note that the convergence radius of the epsilon-expansion for the levels
seems to be smaller than that of the direct epsilon-expansion for certain observables
like critical exponents and impurity susceptibility and entropy,
see Ref.~\onlinecite{MVLF}.


\section{Conclusions}
\label{sec:conclusions}

Using the quantum phase transitions in the soft-gap Anderson model
as an example, we have demonstrated that epsilon-expansion techniques
can be used to determine complete many-body spectra at quantum critical
points.
To this end, we have connected information from standard perturbative RG,
which yields information on critical dimensions and parametrically
small couplings, and from NRG for the many-body spectra of free-fermion
fixed points.
Together, these can be used to perform renormalized perturbation theory
for many-body spectra of {\em interacting} intermediate-coupling fixed
points.
For the soft-gap Anderson model, which features two lower-critical dimensions
at $r=0$ and $r=1/2$, correspondingly two different approaches can be utilized
to capture the same critical fixed point:
Near $r=0$ a Kondo term has to be added to a free-fermion chain with a
decoupled impurity,
whereas near $r=1/2$ an on-site repulsion is used as a perturbation
to the non-interacting Anderson (or resonant-level) model.
These perturbations lift the large degeneracies present in the
non-interacting spectra, and accurately reproduce the critical
spectra determined in NRG calculations at criticality.

Vice versa, our method will be useful in situations where the
effective low-energy theory for the critical point is not known:
a careful analysis of the many-body spectrum near critical dimensions
yields information about the scaling dimension and structure of the
relevant operators.

For instance, a plot similar to Fig.~\ref{fig1} can be calculated for the
spin-boson model, using the numerical renormalization group
method as in Ref.~\onlinecite{BTV}. Preliminary
results (not shown here) indicate that the many-particle levels of the QCP
approach the levels of the delocalized (localized) fixed point in the
limit $s\to0$ ($s\to1$), with $s$ the exponent of the bath spectral
function $J(\omega)\propto\omega^s$.

We envision applications of our ideas to more complex impurity models,
e.g., with two orbitals or two coupled spins, as well as to non-equilibrium
situations treated using NRG \cite{fba}.


\acknowledgments

We thank S. Kehrein, Th. Pruschke, and A. Rosch for discussions
and S. Florens, L. Fritz, M. Kir\'{c}an, and N. Tong for collaborations
on related work.
This research was supported by the DFG through SFB 484 (HJL, RB) and
the Center for Functional Nanostructures Karlsruhe (MV).
MV also acknowledges support from
the Helmholtz Virtual Quantum Phase Transitions Institute in Karlsruhe.


\appendix
\section{Details of the Perturbative Analysis around the Local Moment Fixed Point}
\label{app:A}

In this Appendix, we want to give more details for the derivation
of the matrix $W_{ij}$ eq.~(\ref{eq:Wij}) which determines the
splitting of the fourfold degeneracy of the subspace
($Q=0$, $S=1/2$, $S_z=1/2$) of the LM fixed point at energy
$\epsilon_1 + \epsilon_3$.
We focus on the matrix element $W_{12}$:
\begin{equation}
   W_{12} = \langle \psi_1\vert  H_N^\prime \vert \psi_2 \rangle
          = \alpha(r) f(N)\,  \langle \psi_1\vert
                \vec{S}_{\rm imp} \cdot \vec{s}_0 \vert \psi_2 \rangle
    \ .
\end{equation}
The strategy for the calculations can be extended to the other matrix elements and the other subspaces, for which
we add the perturbative results at the end of this appendix without derivation. The operator $ \vec{S}_{\rm imp} \cdot \vec{s}_0$
is decomposed in four parts:
\begin{eqnarray}
     \vec{S}_{\rm imp} \cdot \vec{s}_0 &=&
      \frac{1}{2}  {S}_{\rm imp}^+ c_{0\downarrow}^\dagger c_{0\uparrow}
   +   \frac{1}{2}  {S}_{\rm imp}^- c_{0\uparrow}^\dagger c_{0\downarrow}
    \nonumber \\
   & & +  \frac{1}{2} {S}_{\rm imp}^z
       \left( c_{0\uparrow}^\dagger c_{0\uparrow} -
              c_{0\downarrow}^\dagger c_{0\downarrow} \right) \ ,
\end{eqnarray}
so that $W_{12}$ can be written as
\begin{equation}
    W_{12} = \alpha(r) f(N)\,  \frac{1}{2}\left[
             {\rm I} + {\rm II} + {\rm III} - {\rm IV}
             \right] \ ,
\end{equation}
with
\begin{equation}
     {\rm I} = \langle \psi_1\vert
                {S}_{\rm imp}^+ c_{0\downarrow}^\dagger c_{0\uparrow}
                 \vert \psi_2 \rangle \ ,
\end{equation}
and the other terms accordingly. With the definitions of
$\vert \psi_1 \rangle$ and $\vert \psi_2 \rangle$ of
eq.~(\ref{eq:four-psis}) we have:
\begin{eqnarray}
   {\rm I} &=& \frac{1}{\sqrt{2}} \langle \psi_0\vert
               \left( \xi^\dagger_{-3\uparrow} \xi_{1\uparrow} +
               \xi^\dagger_{-3\downarrow} \xi_{1\downarrow}
               \right) f_{\uparrow}
               {S}_{\rm imp}^+ c_{0\downarrow}^\dagger c_{0\uparrow}
   \nonumber \\
   & &  \times
       \left[
        \frac{1}{\sqrt{6}} f^\dagger_{\uparrow}
        \left( \xi^\dagger_{1\uparrow} \xi_{-3\uparrow} -
               \xi^\dagger_{1\downarrow} \xi_{-3\downarrow}
        \right) +
        \frac{2}{\sqrt{6}} f^\dagger_{\downarrow}
        \xi^\dagger_{1\uparrow} \xi_{-3\downarrow}
        \right] \vert \psi_0 \rangle  \ .
 \nonumber \\
\end{eqnarray}
With
${S}_{\rm imp}^+ = f^\dagger_{\uparrow} f_{\downarrow}$
we immediately see that the terms containing
$f_{\uparrow}{S}_{\rm imp}^+f^\dagger_{\uparrow}$ drop out.
The remaining impurity operators,
$f_{\uparrow}{S}_{\rm imp}^+f^\dagger_{\downarrow}$, give
unity when acting on  $\vert \psi_0 \rangle$ so one arrives
at
\begin{equation}
   {\rm I} =   \frac{1}{\sqrt{3}} \left[
              {\rm Ia} +  {\rm Ib}
              \right] \ ,
\end{equation}
with
\begin{eqnarray}
    {\rm Ia} &=& \langle \psi_0\vert
     \xi^\dagger_{-3\uparrow} \xi_{1\uparrow}
     c_{0\downarrow}^\dagger c_{0\uparrow}
     \xi^\dagger_{1\uparrow} \xi_{-3\downarrow}  \vert \psi_0 \rangle
    \nonumber \\
    {\rm Ib} &=& \langle \psi_0\vert
     \xi^\dagger_{-3\downarrow} \xi_{1\downarrow}
     c_{0\downarrow}^\dagger c_{0\uparrow}
     \xi^\dagger_{1\uparrow} \xi_{-3\downarrow}  \vert \psi_0 \rangle
    \ .
\end{eqnarray}
To analyze Ia and Ib, the operators $c_{0\sigma}^{(\dagger)}$
have to be expressed in terms of the operators
$\xi^{(\dagger)}_{p\sigma}$:
\begin{equation}
   c_{0\sigma} = \sum_{p^\prime} \alpha_{0p^\prime}
       \xi_{p^\prime\sigma} \ \ , \ \
   c^\dagger_{0\sigma} = \sum_{p} \alpha_{0p}^\ast
       \xi^\dagger_{p\sigma}\ ,
\label{eq:c-xi}
\end{equation}
with the sums over $p$ and $p^\prime$ defined in eq.~(\ref{eq:sum_p}).
This gives
\begin{equation}
 {\rm Ia} = \sum_{pp^\prime} \alpha_{0p}^\ast \alpha_{0p^\prime}
      \langle \psi_0\vert
     \xi^\dagger_{-3\uparrow} \xi_{1\uparrow}
     \xi^\dagger_{p\downarrow}\xi_{p^\prime\uparrow}
     \xi^\dagger_{1\uparrow} \xi_{-3\downarrow}  \vert \psi_0 \rangle
 \ .
\label{eq:Ia}
\end{equation}
The only non-zero matrix elements of eq.~(\ref{eq:Ia}) are for
$p=p^\prime=-3$:
\begin{eqnarray}
 {\rm Ia} &=& \alpha_{0-3}^\ast \alpha_{0-3}
      \langle \psi_0\vert
     \xi^\dagger_{-3\uparrow} \xi_{1\uparrow}
     \xi^\dagger_{-3\downarrow}\xi_{-3\uparrow}
     \xi^\dagger_{1\uparrow} \xi_{-3\downarrow}  \vert \psi_0 \rangle
      \nonumber \\
      &=& - \vert \alpha_{0-3} \vert^2 \ .
\end{eqnarray}
Similarly, the term Ib gives
\begin{eqnarray}
 {\rm Ib} &=& \sum_{pp^\prime} \alpha_{0p}^\ast \alpha_{0p^\prime}
      \langle \psi_0\vert
     \xi^\dagger_{-3\downarrow} \xi_{1\downarrow}
     \xi^\dagger_{p\downarrow}\xi_{p^\prime\uparrow}
     \xi^\dagger_{1\uparrow} \xi_{-3\downarrow}  \vert \psi_0 \rangle
      \nonumber \\
      &=&  \vert \alpha_{01} \vert^2 \ ,
\end{eqnarray}
so that in total:
\begin{equation}
   {\rm I} =   \frac{1}{\sqrt{3}} \left[
             -\vert \alpha_{0-3} \vert^2 + \vert \alpha_{01} \vert^2
              \right] \ .
\end{equation}
The next term
${\rm II} = \langle \psi_1\vert
   {S}_{\rm imp}^- c_{0\uparrow}^\dagger c_{0\downarrow}
   \vert \psi_2 \rangle$ gives
zero due to the combination of impurity operators:
$f_{\uparrow}f^\dagger_{\downarrow}f_{\uparrow}\ldots$
with $f_{\uparrow}$ from $ \langle \psi_1\vert$ and
$f^\dagger_{\downarrow}f_{\uparrow}={S}_{\rm imp}^-$.

The third term
${\rm III} = \langle \psi_1\vert
   {S}_{\rm imp}^z c_{0\uparrow}^\dagger c_{0\uparrow}
   \vert \psi_2 \rangle$
gives
\begin{eqnarray}
  {\rm III} &=& \frac{1}{\sqrt{12}}
    \langle \psi_0\vert
               \left( \xi^\dagger_{-3\uparrow} \xi_{1\uparrow} +
               \xi^\dagger_{-3\downarrow} \xi_{1\downarrow}
               \right) f_{\uparrow}
               {S}_{\rm imp}^z c_{0\uparrow}^\dagger c_{0\uparrow}
               f^\dagger_{\uparrow}
   \nonumber \\
   & &
        \times
        \left( \xi^\dagger_{1\uparrow} \xi_{-3\uparrow} -
               \xi^\dagger_{1\downarrow} \xi_{-3\downarrow}
        \right)\vert \psi_0 \rangle  \ ,
\end{eqnarray}
where the term with
$\frac{2}{\sqrt{6}} f^\dagger_{\downarrow}
        \xi^\dagger_{1\uparrow} \xi_{-3\downarrow}
$
from $\vert \psi_2 \rangle$ has already been dropped.
So we are left with four terms
\begin{equation}
   {\rm III} =  \frac{1}{\sqrt{12}} \left[
      {\rm IIIa} - {\rm IIIb} + {\rm IIIc} - {\rm IIId} \right] \ ,
\end{equation}
with
\begin{eqnarray}
    {\rm IIIa} &=&  \langle \psi_0\vert
     \xi^\dagger_{-3\uparrow} \xi_{1\uparrow}
     f_{\uparrow} {S}_{\rm imp}^z
     c_{0\uparrow}^\dagger c_{0\uparrow}
     f^\dagger_{\uparrow}
     \xi^\dagger_{1\uparrow} \xi_{-3\uparrow}  \vert \psi_0 \rangle \ ,
    \nonumber \\
    {\rm IIIb} &=&\langle \psi_0\vert
     \xi^\dagger_{-3\uparrow} \xi_{1\uparrow}
     f_{\uparrow} {S}_{\rm imp}^z
     c_{0\uparrow}^\dagger c_{0\uparrow}
     f^\dagger_{\uparrow}
     \xi^\dagger_{1\downarrow} \xi_{-3\downarrow}  \vert \psi_0 \rangle \ ,
    \nonumber \\
    {\rm IIIc} &=&\langle \psi_0\vert
     \xi^\dagger_{-3\downarrow} \xi_{1\downarrow}
     f_{\uparrow} {S}_{\rm imp}^z
     c_{0\uparrow}^\dagger c_{0\uparrow}
     f^\dagger_{\uparrow}
     \xi^\dagger_{1\uparrow} \xi_{-3\uparrow}  \vert \psi_0 \rangle \ ,
    \nonumber \\
    {\rm IIId} &=&\langle \psi_0\vert
     \xi^\dagger_{-3\downarrow} \xi_{1\downarrow}
     f_{\uparrow} {S}_{\rm imp}^z
     c_{0\uparrow}^\dagger c_{0\uparrow}
     f^\dagger_{\uparrow}
     \xi^\dagger_{1\downarrow} \xi_{-3\downarrow}  \vert \psi_0 \rangle \ .
    \nonumber \\
\end{eqnarray}
Following similar arguments as above one obtains
\begin{equation}
   {\rm IIIa} =  \frac{1}{2}{\sum_p}^\prime \vert \alpha_{0p} \vert^2 \ ,
\end{equation}
where the $p$ in ${\sum_p}^\prime$ takes the values
\begin{displaymath}
p=1,-1,-5,-7,\ldots -N \ ,
\end{displaymath}
then
\begin{equation}
   {\rm IIIb} = {\rm IIIc} = 0 \ ,
\end{equation}
and
\begin{equation}
   {\rm IIId} =  \frac{1}{2}
{\sum_p}^{\prime\prime} \vert \alpha_{0p} \vert^2 \ ,
\end{equation}
where the $p$ in ${\sum_p}^{\prime\prime}$ takes the values
\begin{displaymath}
p=-1,-3,-5,-7,\ldots -N \ .
\end{displaymath}
This gives for the third term
\begin{eqnarray}
    {\rm III} &=&  \frac{1}{\sqrt{12}} \left[
                   {\rm IIIa}  - {\rm IIId} \right] \nonumber \\
     &=&   \frac{1}{4\sqrt{3}}
           \left[ {\sum_p}^{\prime} \vert \alpha_{0p} \vert^2 -
               {\sum_p}^{\prime\prime} \vert \alpha_{0p} \vert^2 \right]
             \nonumber \\
     &=&    \frac{1}{4\sqrt{3}}
           \left[ \vert \alpha_{01} \vert^2 -
                    \vert \alpha_{0-3} \vert^2\right] \ .
\end{eqnarray}
The calculation of IV proceeds very similarly to III and one obtains
\begin{equation}
   {\rm III} = - {\rm IV} \ ,
\end{equation}
so that we finally arrive at
\begin{eqnarray}
   W_{12} &=&  \alpha(r) f(N) \frac{1}{2}
             \left( \vert \alpha_{01} \vert^2 -
                    \vert \alpha_{0-3} \vert^2
             \right)
             \left[ \frac{1}{\sqrt{3}} + 0 + 2  \frac{1}{4\sqrt{3}} \right]
 \nonumber \\
 &=&  \alpha(r) f(N) \frac{1}{4} \sqrt{3} \left( \vert \alpha_{01} \vert^2 -
                    \vert \alpha_{0-3} \vert^2
             \right) \ .
\end{eqnarray}

We performed a similar analysis for a couple of other subspaces.
Here we list the results from the perturbative analysis for three
more subspaces together with the corresponding basis states.

\subsection{$Q=0$, $S=1/2$, $S_z=1/2$, $E=2\epsilon_1$}
\label{sec:sub2}

This subspace has the same quantum numbers $Q$, $S$ and $S_z$
as the one discussed above, so that the details of the calculation
are very similar.
The differences  originate from the position of particles and holes in the
single-particle spectrum of Fig. \ref{fig:lmfp-spl}. This
reduces the dimensionality of the subspace from four to two.

The corresponding basis can be written as
\begin{eqnarray}
 |\psi_1\rangle&=&
\frac{1}{\sqrt{2}}f_{\uparrow}^{\dagger}
(\xi_{1\uparrow}^{\dagger}\xi_{-1\uparrow}+\xi_{1\downarrow}^{\dagger}\xi_{-1\downarrow})
|\psi_0\rangle \ , \nonumber\\
 |\psi_2\rangle&=&
\left[\frac{1}{\sqrt{6}}f_{\uparrow}^{\dagger}
(\xi_{1\uparrow}^{\dagger}\xi_{-1\uparrow}-\xi_{1\downarrow}^{\dagger}\xi_{-1\downarrow})
+\frac{2}{\sqrt{6}}f_{\downarrow}^{\dagger}\xi_{1\uparrow}^{\dagger}\xi_{-1\downarrow}\right]
|\psi_0\rangle   \ . \nonumber\\
\end{eqnarray}
The first-order corrections are given by the 2$\times$2 matrix
\begin{equation}
\{W_{ij}\}=\alpha(r) f(N)\left[
\begin{array} {cc}
  0 & \frac{\sqrt{3}}{4}\gamma\\
\frac{\sqrt{3}}{4}\gamma & -\frac{1}{2}\beta
\end{array}
\right] \ ,
\end{equation}
with $\gamma=|\alpha_{01}|^2-|\alpha_{0-1}|^2$ and $\beta=|\alpha_{01}|^2+|\alpha_{0-1}|^2$.
Due to the particle-hole symmetry of the conduction band we have
$|\alpha_{01}|=|\alpha_{0-1}|$; therefore, the off-diagonal matrix elements
vanish and the effect of the perturbation is simply a negative energy-shift
only for the state $|\psi_2\rangle$:
\begin{equation}
\{W_{ij}\}=\alpha(r) f(N)\left[
\begin{array} {cc}
  0 & 0\\
0& -|\alpha_{01}|^2
\end{array}
\right] \ .
\end{equation}
This effect can be seen in the energy splitting
of the first two low-lying excitations in Fig. \ref{fig:compare_LM}.

\subsection{$Q=-1$, $S=0$, $E=-\epsilon_{-1}$}
There is only one configuration for this combination
of quantum numbers and excitation energy:
\begin{equation}
 |\psi\rangle=\frac{1}{\sqrt{2}}(f_{\uparrow}^{\dagger}\xi_{-1\uparrow}+f_{\downarrow}^{\dagger}\xi_{-1\downarrow})|\psi_0\rangle \ .
\end{equation}
The first-order perturbation keeps the state in this one-dimensional subspace
 and the energy correction is given by
\begin{equation}
\Delta E=\langle\psi|H_N^\prime|\psi\rangle=-\frac{3}{4}\alpha(r) f(N)|\alpha_{0-1}|^2 \ .
\end{equation}

\subsection{$Q=-1$, $S=0$, $E=-\epsilon_{-3}$}
The difference to the previous case is the position of the hole in the single-particle
spectrum. The state is now given by
\begin{equation}
 |\psi\rangle=\frac{1}{\sqrt{2}}(f_{\uparrow}^{\dagger}\xi_{-3\uparrow}+f_{\downarrow}^{\dagger}\xi_{-3\downarrow})|\psi_0\rangle \ ,
\end{equation}
with the energy correction
\begin{equation}
\Delta E=\langle\psi|H_N^\prime|\psi\rangle=-\frac{3}{4}\alpha(r) f(N)|\alpha_{0-3}|^2 \ .
\end{equation}


\section{Details of the Perturbative Analysis around the Strong Coupling Fixed Point}
\label{app:B}
The main difference in the calculation of the matrix elements $\{W_{ij}\}$ for this
case is due to the structure of the perturbation, see eq.~(\ref{eq:pertU}). Furthermore
the ground state of the SC fixed point is fourfold degenerate and the
perturbation partially splits this degeneracy, as discussed in the following.

\subsection{$Q=0$, $S=1/2$, $S_z=1/2$, $E=0$}
This is one of the four degenerate ground states at the SC fixed point:
\begin{equation}
 |\psi_{1}\rangle=\xi_{0\uparrow}^{\dagger}|\psi_0\rangle \ ,
\label{eq:app-psi0}
\end{equation}
with $|\psi_0\rangle$ defined in eq. (\ref{eq:psi0_U}).

The perturbative correction is given by
\begin{equation}
\langle \psi_1|H_N^{\prime}|\psi_1\rangle=\frac{1}{2}\beta(r) \bar{f}(N)(1-|\alpha_{f0}|^4)
\end{equation}
which corresponds to the energy shift of the ground state:
\begin{equation}
\Delta E_1=\frac{1}{2}\beta(r) \bar{f}(N)(1-|\alpha_{f0}|^4) \ .
\end{equation}
The coefficients $\alpha_{fl}$ are defined via the
relation between the operators
$f_{\sigma}^{(\dagger)}$
and
$\xi^{(\dagger)}_{l\sigma}$:
\begin{equation}
   f_{\sigma} = \sum_{l^\prime} \alpha_{fl^\prime}
       \xi_{l^\prime\sigma} \ \ , \ \
   f^\dagger_{\sigma} = \sum_{l} \alpha_{fl}^\ast
       \xi^\dagger_{l\sigma}\ .
\label{eq:f-xi}
\end{equation}

\subsection{$Q=-1$, $S=0$, $E=0$}
\label{sec:subx1}

This state is also a ground state in the $U=0$ case:
\begin{equation}
 |\psi_{2}\rangle=|\psi_0\rangle \ .
\end{equation}
The calculation of the first-order correction for $|\psi_2\rangle$ gives
\begin{equation}
\langle \psi_2|H_N^{\prime}|\psi_2\rangle=\frac{1}{2}\beta(r) \bar{f}(N)(1+|\alpha_{f0}|^4) \ .
\label{eq:app-deltaE}
\end{equation}
This means that the ground state including the effect of the perturbation
is given by $|\psi_{1}\rangle$ in eq. (\ref{eq:app-psi0}) and the state
$|\psi_{2}\rangle$ appears as an excited state.
For a comparison with the energy levels shown in the NRG flow diagrams,
where the ground state energy is set to zero in each iteration,
we subtract the perturbative correction of the ground state ($\Delta E_1$)
from the energies of all other excited states.
Subtracting this energy shift from eq. (\ref{eq:app-deltaE})
gives the net energy correction for the $|\psi_2\rangle$ state:
\begin{equation}
\Delta E_2=\beta(r) \bar{f}(N)|\alpha_{f0}|^4 \ .
\end{equation}

\vspace*{0.5cm}

\subsection{$Q=-1$, $S=0$, $E=\epsilon_2$}
\label{sec:subx2}
The state corresponding to this subspace is given by:
\begin{equation}
 |\psi_{3}\rangle=\frac{1}{\sqrt{2}}(\xi^{\dagger}_{0\uparrow}\xi_{-2\uparrow}+\xi^{\dagger}_{0\downarrow}\xi_{-2\downarrow})|\psi_0\rangle \ .
\end{equation}
The first-order correction reads
\begin{equation}
\langle \psi_3|H_N^{\prime}|\psi_3\rangle =\beta(r) \bar{f}(N)\left[ \frac{1}{2}(1-|\alpha_{f0}|^4)+3|\alpha_{f0}|^2|\alpha_{f-2}|^2 \right] \ .
\end{equation}
Subtracting the energy correction for the ground state results in
\begin{equation}
\Delta E_3=3\beta(r) \bar{f}(N)|\alpha_{f0}|^2|\alpha_{f-2}|^2 \ .
\end{equation}

\subsection{$Q=-1$, $S=0$, $E=\epsilon_4$}
\label{sec:subx3}
Similarly for the state
\begin{equation}
 |\psi_{4}\rangle=\frac{1}{\sqrt{2}}(\xi^{\dagger}_{0\uparrow}\xi_{-4\uparrow}+\xi^{\dagger}_{0\downarrow}\xi_{-4\downarrow})|\psi_0\rangle \ ,
\end{equation}
the first-order correction is given by
\begin{equation}
\langle \psi_4|H_N^{\prime}|\psi_4\rangle \nonumber \\
 =\beta(r) \bar{f}(N)\left[\frac{1}{2}(1-|\alpha_{f0}|^4)+
3|\alpha_{f0}|^2|\alpha_{f-4}|^2 \right] \ ,
\end{equation}
and subtracting the energy correction for the ground state results in:
\begin{equation}
\Delta E_4=3\beta(r) \bar{f}(N)|\alpha_{f0}|^2|\alpha_{f-4}|^2 \ .
\end{equation}


\end{document}